\documentclass[letterpaper,twocolumn,10pt]{article}
\usepackage{usenix2019_v3}

\usepackage{tikz}
\usepackage{amsmath}
\usepackage{balance}

\input{section/preamble/preamble}
\input{section/preamble/preamblelistingsconf}
\acrodef{CPS}{Cyber-Physical System}
\acrodef{IoT}{Internet of Things}
\acrodef{HDL}{Hardware Description Language}
\acrodef{CAD}{Computer-Aided Design}
\acrodef{EDA}{Electronic Design Automation}
\acrodef{HPC}{High-Performance Computing}
\acrodef{DL}{deep learning}
\acrodef{ML}{machine learning}
\acrodef{NLP}{natural language processing}
\acrodef{IC}{Integrated Circuit}
\acrodef{CWE}[CWE]{Common Weakness Enumeration}
\acrodef{CVE}[CVE]{Common Vulnerabilities and Exposures}
\acrodef{LLM}[LLM]{large language model}
\acrodef{NMT}[NMT]{neural machine translation}
\newcommand{\ignore}[1]{{}}

\newcommand{\squishlist}{
	\begin{list}{$\bullet$}
		{ \setlength{\itemsep}{0pt}
			\setlength{\parsep}{1pt}
			\setlength{\topsep}{1pt}
			\setlength{\partopsep}{0pt}
			\setlength{\leftmargin}{0.9em}
			\setlength{\labelwidth}{1.5em}
			\setlength{\labelsep}{0.4em} } }
	\newcommand{\squishend}{
	\end{list}  }

\definecolor{graphFirst}{RGB}{2,136,209} %
\definecolor{graphSecond}{RGB}{211,47,47} %
\definecolor{graphThird}{RGB}{245,124,0} %
\definecolor{graphFourth}{RGB}{56,142,60} %
\definecolor{graphFifth}{RGB}{81,45,168} %
\definecolor{graphSixth}{RGB}{69,90,100} %
\definecolor{graphSeventh}{RGB}{251,192,45} %
\definecolor{backgroundSecond}{RGB}{239,154,154} %
\definecolor{backgroundThird}{RGB}{255,204,128} %
\definecolor{backgroundFourth}{RGB}{165,214,167} %
\definecolor{backgroundFifth}{RGB}{179,157,219} %
\definecolor{backgroundSixth}{RGB}{176,190,197} %
\definecolor{backgroundSeventh}{RGB}{255,245,157} %

\begin{document}

\date{}

\title{\Large \bf FLAG: Finding Line Anomalies (in code) with Generative AI}

\author{
{\rm Baleegh Ahmad}\\
New York University
\and
{\rm Benjamin Tan}\\
University of Calgary
\and
{\rm Ramesh Karri}\\
New York University
\and
{\rm Hammond Pearce}\\
University of New South Wales
} %

\maketitle

\begin{abstract}
Code contains security and functional bugs. The process of identifying and localizing them is difficult and relies on human labor.
In this work, we present a novel approach (FLAG) to assist human debuggers.
FLAG is based on the lexical capabilities of generative AI, specifically, Large Language Models (LLMs).
Here, we input a code file then extract and regenerate each line within that file for self-comparison. 
By comparing the original code with an LLM-generated alternative, we can flag notable differences as anomalies for further inspection, with features such as distance from comments and LLM confidence also aiding this classification.
This reduces the inspection search space for the designer.
Unlike other automated approaches in this area, FLAG is language-agnostic, can work on incomplete (and even non-compiling) code and requires no creation of security properties, functional tests or definition of rules.
In this work, we explore the features that help LLMs in this classification and evaluate the performance of FLAG on known bugs.
We use  121 benchmarks across C, Python and Verilog; with each benchmark containing a known security or functional weakness. We conduct the experiments using two state of the art LLMs in OpenAI's code-davinci-002 and gpt-3.5-turbo, but our approach may be used by other models. 
FLAG can identify 101 of the defects and helps reduce the search space to $12-17\%$ of source code.

\end{abstract}

\section{Introduction\label{sec:intro}}
Bugs occur in code when there is a discrepancy between a developer's intent and their implementation.
These bugs can introduce security vulnerabilities or functional deficiencies.
Finding them is a laborious process---though assisted tools exist, they typically will only work when programs are fully completed (or at least compilable), and will focus only on a subset of languages or bug classes.
Hence, during the development process, developers and their teams should regularly check their work, and so approaches will combine manual review with automated testing~\cite{berner_observations_2005}, static analysis~\cite{chess_static_2004}, and fuzzing~\cite{li_fuzzing_2018} to help identify potential problems.  
Code must be checked against intentions that are captured intrinsically, explicitly in the case of artifacts like tests, assertions, or rules (e.g., queries when using CodeQL~\cite{github_inc_codeql_2021}) or implicitly (when executing code for crashes in fuzzing).

Given that instances of buggy code are infrequent relative to correct code (average
industry code has been estimated as containing 0.5 and 25 bugs per 1,000 lines~\cite{mcconnell_code_2004}), we assume that developer intent is mostly captured in source code and comments with occasional lapses that need to be found and dealt with. 
Imagine a developer who wants to review their code manually; if a program comprises only a few lines of code and comments, it is probably feasible for both novices and experts. 
As projects grow, this becomes increasingly challenging by virtue of scale. 
Finding ways to narrow down potential problem areas in code for review can help developers, especially under a time crunch. 
\textit{Can the intent of the source code and comments be used to flag problems in the code, narrowing the extent of manual review needed? } To answer this, we investigate the use of Generative AI, specifically large language models (LLMs), to FLAG  anomalies in code.

LLMs such as GPT-3~\cite{brown_language_2020} and Codex~\cite{chen_evaluating_2021} demonstrate significant capability for number of lexical tasks including code-writing. They produce outputs based on the continuation of their inputs---a kind of `smart autocomplete'.
These inputs can be existing code and comments, and the models will produce probablistically likely matching code.  This raises an interesting possibility: if (a) buggy lines of code are a minority of those present, and (b) a majority of the code aligns with the intent of the author, is it plausible to use LLMs to measure if a given line of code is an outlier? If so, this is a strong indication that the line was unusual and potentially defective.

Using this intuition, we propose a novel approach where we use LLMs to generate alternative lines of code given existing code and comments; these alternatives are compared against the developer's code to identify potential discrepancies. 
Recent work has motivated the exploration of LLMs, both in terms of their impact on security (e.g., in vulnerability introduction~\cite{pearce_asleep_2022} and user studies~\cite{sandoval_lost_2023}) and their use for improving security (e.g., bug repair~\cite{pearce_examining_2022}). 
This work provides complementary insights into whether  LLMs can be used to help identify bugs both in the form of security vulnerabilities and functional deficiencies. 
Our \textbf{contributions} are as follows. 
\begin{itemize}
    \item We propose FLAG, a framework for the novel use of LLMs in the detection of bugs by comparing original code with LLM-generated code. The details of the techniques used in the tool are mentioned in~\autoref{sec:flag}.
    \item We explore features of source code and information from LLMs to classify code as buggy or not. Effectiveness of these features is analyzed by performing experiments for major LLMs in various modes across multiple languages in~\autoref{subsec:tool-experiments-and-results}, with further discussion in~\autoref{sec:discussion}.
    \item The tool and results are open-sourced  at~\cite{review_artifacts_2023}.
\end{itemize}

\section{Background\label{sec:background}}

In this section we discuss bug detection, how LLMs work and the information available in source code that LLMs can use. This conveys the motivation for why FLAG is needed before discussing the details of its implementation in~\autoref{sec:flag}.

\subsection{Large Language Models (LLMs)}

LLMs (including GPT-2~\cite{radford_language_2019}, GPT-3~\cite{brown_language_2020}, and Codex~\cite{chen_evaluating_2021}) are based on the Transformer~\cite{vaswani_attention_2017} architecture. 
They can be thought of as ``scalable sequence prediction models''~\cite{chen_evaluating_2021} capable of a wide range of lexical tasks.
When provided with a prompt consisting of a sequence of tokens, they generate the most probable set of tokens to continue or complete the sequence, similar to an intelligent autocomplete feature. 
In this context, tokens refer to common sequences of around four characters long and are assigned a unique identifier within a user-defined vocabulary size. This byte pair encoding (BPE)~\cite{gage_new_1994} enables the LLMs to process a larger amount of text within their fixed-size input window. Thus most LLMs work over tokens rather than individual characters.

Once trained over suitable examples, an LLM can be used to fill in the body of a function based on its signature and/or comment~\cite{chen_evaluating_2021}. In the case of the commercial models we choose to investigate, this corpus is made up of billions of lines of open-source code scraped from the internet (e.g. GitHub).

LLMs range in size and capabilities. In this work we evaluate 2 OpenAI LLMs, code-davinci-002 and gpt-3.5-turbo. code-davinci-002 is a Codex model that is optimized for code completion tasks. We use it in two completion modes, without and with suffixes, referred to as auto-complete and insertion, respectively. gpt-3.5-turbo improves on GPT-3 and can understand as well as generate natural language or code. We use it in two modes, without any instruction or with an instruction to generate the next line of code, referred to as auto-complete and instructed-complete respectively.

\subsection{Role of comments}

Comments in code are often ignored for static code analysis. This is because the quality of comments is highly variable, and they do not play a role in the actual functionality of the program. They are, however, a good source of documentation for the intended behavior of the program~\cite{steidl_quality_2013}. Just like a human uses comments to reason about the code, tools like LLMs also possess the same ability.

Some efforts have utilized comments in an attempt to detect weaknesses in code. The most relevant work is icomment~\cite{tan_icomment_2007} which utilizes Natural Language Processing to automatically analyze comments and detect inconsistencies between comments and source code. They reason that these inconsistencies could contain a bug because the comment was correct but corresponding code implementation was wrong. On the other hand, the code could be correct, but the comment was bad. icomment takes comments and forms rules for source code to pass. Failure of these rules are reported as inconsistencies.
Another relevant work is @tComment \cite{tan_tcomment_2012} which focuses on Javadoc comments. They use the same insight of code comment inconsistencies to take source code files for Java to infer properties for methods and then generate random tests for these methods. Failure of these tests are reported as inconsistencies.
FLAG takes the insight of comment code inconsistencies but instead of generating rules or tests, generates alternate code for comparison with original code. Moreover, FLAG uses previous code plus comments for its operation.

\subsection{Bug detection}
Static detectors are used in various shapes and forms across many software companies. They are typically used in the development stage before deployment to catch bugs in code. Google's Error Prone \cite{error_prone_error_2023} catches common programming mistakes in Java. Facebook's Infer \cite{infer_infer_2021} does the same for Java and C/C++/Objective C code. And other well-known tools include SpotBugs \cite{spotbugs_spotbugs_2022} and CodeQL \cite{github_inc_codeql_2021}.
They typically work by conducting analysis over the Abstract Syntax Tree (AST) and/or data-flow graph of the code \cite{habib_how_2018}. These constructs are traversed with an elaborate set of checkers that are used to indicate improper behavior of code e.g., CodeQL could be used to create a query that checks whether a path exists between two nodes that should not otherwise. This improper behavior could be a pattern in the AST or some flow in the data-flow graph. 
Detectors may also infer rules from version histories and source code comments.
While static detectors are generalizable across databases of the same language, they require the creation of a large set of patterns and known buggy flows for detecting bugs. Moreover, they are only able to find bugs in this limited knowledge base of patterns and flows.

Unit testing is the other methodology commonly used in the struggle to identify defects in code \cite{randoop_randoop_2023, evosuite_release_2021}. The obvious challenge with unit tests is the strict requirement for knowledge of the functionality of the program. The workaround employed is the development of automatically generated unit tests. Code coverage through this approach remains limited, and even when the coverage is there, the faults are sometimes not revealed \cite{shamshiri_automatically_2015}. Additionally, unit tests do not cater to affirming the security of the code.

\subsection{Machine Learning based Detectors}
Researchers have used language model-based techniques to try and detect bugs. 
Bugram \cite{wang_bugram_2016} uses N-gram Language models to obtain sequences for tokens in programs. These token sequences are explored according to their probabilities in the learned model, and the ones with lower probabilities were marked as probable bugs.
Hoppity \cite{dinella_hoppity_2020} is a learning-based approach relying on graph transformations to detect and fix bugs. The graph model of the source code is used to make a series of predictions regarding the position of bug nodes and corresponding graph edits to produce a fix.
EnSpec \cite{chakraborty_entropy_2018} is a method that uses code entropy (a metric devised to represent the \textit{naturalness} of code derived from a statistical language model) for bug localization. They use the intuition that buggy code tends to be higher in entropy to facilitate localization.
In another work, solutions of bugs and a language model based on long short-term memory (LSTM) networks was used for bug detection \cite{teshima_bug_2018}. The researchers of this work trained their model over Aizu Online Judge (AOJ) \cite{aizu_aizu_2015}, which contains several million lines of source code.

More recently, LLMs have been explored for this purpose. The difference between LLMs and other language models is the size of training data and complexity of the network.
FuzzGPT \cite{deng_large_2023} uses LLMs as edge case fuzzers by priming LLMs to produce unusual programs for fuzzing. First, they allow LLMs to directly learn from historical reported bugs and then generate similar bug-triggering code snippets to find new bugs. They use Codex and CodeGen models to detect bugs in the popular DL libraries PyTorch and TensorFlow.
In another work, Li et al.~\cite{li_finding_2023} show how ChatGPT can be used through differential prompting to detect bugs in the Quixbugs \cite{lin_quixbugs_2017} database. This involves generating reference designs for a problem using ChatGPT. For a given test input, if reference designs produce the same output, but the buggy version produces a different one, the test case is identified as a failure-inducing test case.
DeepBugs \cite{pradel_deepbugs_2018} uses natural language elements in code to implement a name-based bug detection machine learning tool. The key idea is to convert function names and identifiers  into embeddings that are learned by the network in order to preserve semantic information. They applied their approach to a corpus of JavaScript files.
While these efforts are a step forward in using  LLMs to detect bugs, they either target a niche subset of code or require information like test cases for a program. The ability of LLMs to fix bugs is not evaluated in a generalizable way, i.e., using no information or knowledge outside of source code.

\subsection{Why FLAG?}

The limitations mentioned for static detectors and machine learning detectors leave room for a tool that does not require huge efforts to set up security rules and tests and is applicable on many programming languages.
Through our various experiments, we have demonstrated that our checking approach is agnostic to the language of code. Our insights can be used to leverage off-the-shelf LLMs that work on various languages for providing code feedback.
FLAG also does not require code to be compiled or even syntactically correct, enabling it to work on incomplete code. 
This allows vulnerability checking at earlier stages of design when compared to traditional static checkers.
Additionally, FLAG does not require creation of security rules and tests. This is a big plus because security checks are by nature non-exhaustive and require a lot of time and domain expertise to devise. This is particularly true for hardware where formal tools have been shown to fail in detecting a lot of RTL  security bugs~\cite{dessouky_hardfails_2019}.

\begin{figure}[t]
    \centering
    \includegraphics[width=\linewidth]{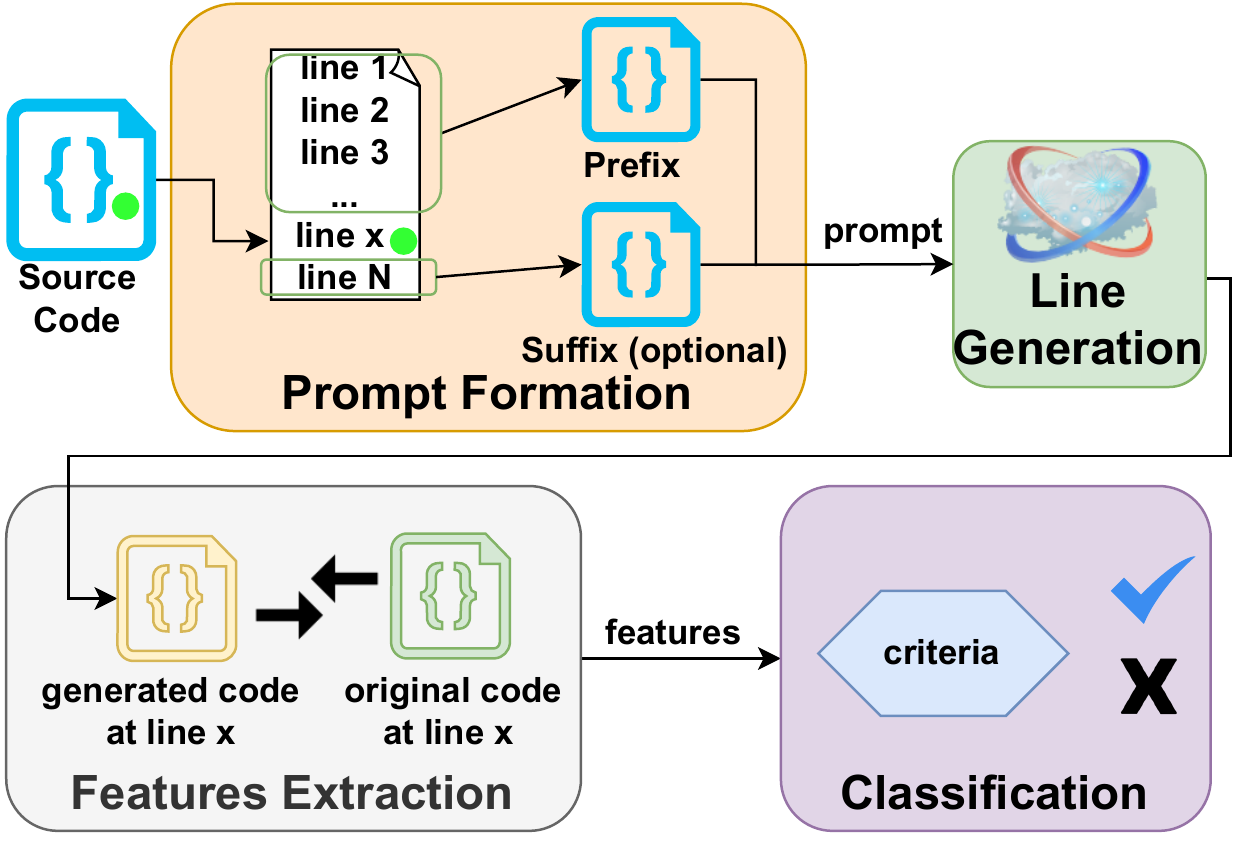}
    \caption{FLAG pipeline for Finding Line Anomalies.}
    \label{fig:tool}
\end{figure}

\section{FLAG Approach%
\label{sec:flag}}

The FLAG consistency checker relies on the unique capabilities of LLMs for writing code. %
It has the approach shown in \autoref{fig:tool}. For each line in the source code, FLAG generates a corresponding prompt which consists of the part of code before the line (prefix) and optionally, the code after the line of code (suffix). The prompt consisting of code and comment is the input to the LLM which outputs a single line of code or comment. %
This generated line is compared to the original line to produce \textit{features} that are used in the classification of the original line as buggy or not. These \textit{features} provide either a quantitative estimate of the difference between the two lines or the degree of confidence of the line generated by the LLM. They are discussed in \autoref{subsec:features-extraction}. The lines classified as potentially buggy are flagged to the designer. The FLAG flow for a given line has a 4-step process of prompt formation, line generation, feature extraction, and classification.

To begin the consistency checks on a file, we must give FLAG a line to start checking from. This is done to give the LLM enough context to start producing relevant code and comments. Normally, we give the starting line number after the header definitions, initial comments, and declaration of modules and internal signals. In some cases, when the defect is part of the declared signals, we make an exception to start from the beginning of the file. 
Additionally, the file is preprocessed to skip  empty lines and to identify if a comment was present before the line to start checking from.

\subsection{Prompt Formation\label{prompt-formation}}

\begin{figure}
\centering
\begin{subfigure}[b]{0.95\linewidth}
\lstset{numbers=left,language=C,breaklines, aboveskip=0pt,belowskip=0pt}

    \lstinputlisting[lastline=11,backgroundcolor=\color{yellow},frame=l]{llm-input-example.c}
    \lstinputlisting[backgroundcolor=\color{pink},firstline=12,lastline=12, firstnumber=12,frame=l]{llm-input-example.c}
    \lstinputlisting[firstline=13, firstnumber=13,backgroundcolor=\color{lightgray},frame=l]{llm-input-example.c}
\caption{Step 1: Traversing  original file for LLM input. Target line:12 (pink). Prefix lines 1-11 ( yellow) and optional suffix lines 13-15 (gray).}
\label{fig:llm-input-example-prefix-and-suffix}
\end{subfigure}

\begin{subfigure}[b]{0.95\linewidth}
\begin{lstlisting}[language=C,frame=l]
if (index >= 0 && index < size) {
\end{lstlisting}
\caption{Step 2: Response by gpt-3.5-turbo in auto-complete mode. This is detected as different to the original line 12, which is thus flagged for inspection, where a defect (CWE-125, out-of-bounds read) is found.}
\label{fig:llm-input-example-output}
\end{subfigure}

\caption{LLM consistency check on line 12 of benchmark C1-10 (CWE-125) in source C1.}
\label{fig:llm-inputs-example}
\end{figure}

Prompt formation takes the source code and line to be classified as the input and produces the prompt as the output. This prompt is  sent to the LLM for line generation. An example  is shown in~\autoref{fig:llm-input-example-prefix-and-suffix}. The file shown is from source C1 for the defect containing CWE-125. The prompt generated for line 12 consists of the prefix covering lines 1-11 and the suffix covering lines 13-15. This process is repeated for each line in the file. As the FLAG checker progresses line by line, the prefix increases while the suffix decreases.

We limit  prefix and suffix lengths to a maximum of 50 lines to stay within the token limits of the LLMs. The prompt is tweaked to elicit a better response from the LLM e.g., if the LLM generates a comment when it should be generating code, we append the first few characters of the original line of code to the prefix. This process may be repeated multiple times till a valid response is obtained and requires feedback from line generation. This is why it is discussed in~\autoref{subsec:line-generation}. The suffix is used only when the model supports insertion mode.

\subsection{Line Generation\label{subsec:line-generation}}

\begin{algorithm}[!tbp]
\caption{LLM line generation}\label{alg:llm-line-generation}
\begin{algorithmic}[1]

\Procedure{generate}{$orig\_lines, loc, num\_lines$ }\\

\textbf{If} $loc>max\_pre\_len$\\

\quad $prefix \leftarrow orig\_lines$[loc-max\_pre\_len:loc] \\

\textbf{Else} $prefix \leftarrow orig\_lines$[:loc]\\

\textbf{If} $num\_lines-loc>max\_suf\_len$\\

\quad $suffix \leftarrow orig\_lines$[loc+1:loc+$max\_suf\_len$] \\

\textbf{Else} $suffix \leftarrow orig\_lines$[loc+1:]\\

$max\_attempts \leftarrow$3, $attempts \leftarrow$0, $generated \leftarrow$False\\

\textbf{while} $generated$==False\\

\quad    \textbf{Try} \\
    
\quad\quad        \textbf{If} $attempts>0$\\
        
\quad\quad\quad        Append $orig\_lines$[loc][:4] to $prefix$\\
        
\quad\quad\quad       $response \leftarrow$ Generation for given $prefix$ and 

\quad\quad\quad\quad\quad      $suffix$ (optional) by LLM\\
        
\quad    \textbf{Except} Note error\\
    
\quad    \textbf{Else} \\
    
\quad\quad        \textbf{If} $response$==`' and $attempts$ < $max\_attempts$:\\
        
\quad\quad\quad        Increment $attempts$\\
        
\quad\quad        \textbf{Elif} $orig\_lines$[loc] is code and $response$ is not \\
        
\quad\quad\quad        Increment $attempts$ \\
        
\quad        \textbf{Else} $generated \leftarrow$True \\

\textbf{end while}
\EndProcedure

\end{algorithmic}
\end{algorithm}

Line generation takes the prompt as an input and outputs a line of code or comment generated by the LLM. For example, the line generated for the prompt in~\autoref{fig:llm-input-example-prefix-and-suffix} is shown in ~\autoref{fig:llm-input-example-output}. In this instance, the response by LLM is different from the original line of code. It is flagged for inspection, revealing that the original line of code contained the security defect CWE-125, i.e., out-of-bounds read. The LLM used for this example was gpt-3.5-turbo in auto-complete mode. Details of experiments are in~\autoref{subsec:tool-experiments-and-results}.

The LLM is guided to produce a reasonable output because sometimes the LLM may return an empty line or return a comment instead of code. This process is described in \autoref{alg:llm-line-generation}. 
\texttt{orig\_lines} is a list of the lines in the original file. Assume that the list is indexed starting 1. \texttt{loc} is the line number that we want the LLM to generate. \texttt{num\_lines} is the total number of lines in the original file (after being preprocessed).
We initialize the prefix and suffix as empty strings and use \texttt{max\_pre\_len} and \texttt{max\_suf\_len} as the limits of their sizes, respectively. This is done as an attempt to keep the token size of the prompt reasonable. For our experiments, we set the limits to 50. The assignment of appropriate content from the original file to prefixes and suffixes is shown in lines 2-7.
To overcome occasional unusable outputs, we prompt the LLM to produce a response again, up to a maximum of \texttt{max\_attempts} times. This is shown in the Try and Catch block in lines 15-19. Lines 17 and 19 increment the number of attempts and return to the start of the Try block at line 10.  On the first try, an LLM is prompted to produce a response with the given prefix and suffix.  On the second and third attempts, we provide assistance to elicit a non-empty response. This is done by appending the first 5 characters of the line the LLM is attempting to generate to the prefix. This assist is shown in line 12. If there is any error in the try block, the error is noted in line 14 before returning to line 10. If there is no error and the conditions in lines 16 and 18 do not hold, line 20 is executed. This asserts the signal that a line has been successfully generated and the while loop is exited. For all completions, we use a temperature of 0, max\_token limit of 150, top\_p value of 1, and end line character as the stop token.

\subsection{Feature Extraction\label{subsec:features-extraction}}

Feature extraction takes the original and buggy lines of code and outputs \textit{features}. These are quantitative values that are used for the classification of the original line as potentially buggy or not. They represent the difference between the two lines or the confidence of the generated code.

\subsubsection{Features \label{subsubsec:features}}

\texttt{Levenshtein Distance (ld)} is an edit distance between two strings. It considers three operations: addition, deletion, and substitution. The sum of the number of operations required to convert one string to another is the Levenshtein Distance. A perfect match results in a score of 0. We use \texttt{ld} to compare lines of code. Since the goal is to flag defects in code, we use \texttt{ld} as the primary criterion to identify defects. 

\texttt{BLEU} i.e., Bilingual Evaluation Understudy Score~\cite{papineni_bleu_2002}, is a metric for evaluating a candidate sentence to a reference sentence. A perfect match results in a score
of 1.0, whereas a perfect mismatch results in a score of 0.0. We use it for comparing comments as they resemble natural language. We collect the cumulative BLEU-1 to BLEU-4 scores but found that only BLEU-1 produced meaningful numbers. BLEU-2, BLEU-3 and BLEU-4 had minuscule quantities, which are not useful. For the rest of this paper, \texttt{BLEU} refers to BLEU-1.

\texttt{Distance from comment (dfc)} indicates how far a line of code is from the closest comment before it. If the line also contains a comment, \texttt{dfc} has the value 0. If there is no comment before the line, \texttt{dfc} has no value. We only consider comments before code as relevant to the code because that is typically how code is written. 

\texttt{logprob} in the context of code generation by LLMs is the log of the probability of the token generated. 
If a token is more likely to be generated, it will have a higher \texttt{logprob} with its maximum possible value being 0. If a token is less likely to be generated, it will have a lower \texttt{logprob}, i.e., a negative value with higher magnitude.
An LLM tends to ``choose'' a token with a higher probability.  For a generated line, we take the average of the logprobs of the tokens generated, which we refer to as \texttt{logprob} from here on. \texttt{logprob} closer to 0 indicates more confidence in the generation, whereas a more negative value indicates lesser confidence.

\subsubsection{Extraction \label{subsubsec:extraction}}

\begin{table*}[]
\centering
\caption{Lines generated by gpt-3.5-turbo (auto-complete mode) and features extracted for benchmark C1-10(CWE-125). Line with true defect is shaded in pink. Lines flagged according to criterion C2(20,10) are marked with a check in \textbf{Flagged?} column. File shown partially for brevity.}
\label{tbl:example-features}
\small
\begin{tabular}{|L{6.5cm}|L{5.5cm}|l|l|l|c|}
\hline
\textbf{Original} & \textbf{Generated} &\textbf{dfc}& \textbf{ld} & \textbf{BLEU} & \textbf{Flagged?}  \\ \hline
printf("\%d\textbackslash{}r\textbackslash{}", getValueFromArray(id\_sequence, 7, id\_index));         & getValueFromArray(id\_sequence, sizeof(id\_sequence)/sizeof(int),id\_index)); & 6   & 67 & --  &  \\ \hline
return 0;                                                                                              & return 0;                                                                     & 7   & 0  & -- &   \\ \hline
\}                                                                                                     & \}                                                                            & 8   & 0  & --  &  \\ \hline
int getValueFromArray (int* array, int size, int index) \{                                             & int getValueFromArray(int* array, int size, int index)\{                      & 10  & 1  & -- &   \\ \hline
//given the array of integers return the value at the given index and -1 if the index is out of bounds & //if the index is out of bounds return -1                                     & 0   &    & 0.425 & \\ \hline
\rowcolor[HTML]{FFCCC9} 
if(index \textless{}size) \{                                                                           & if (index \textgreater{}= 0 \&\& index \textless{}size) \{                    & 1   & 15 & -- &  $\checkmark$ \\ \hline
return array{[}index{]};                                                                               & return array{[}index{]};                                                      & 2   & 0  & --  &  \\ \hline
\}                                                                                                     & \} else \{                                                                    & 3   & 7  & --  & $\checkmark$  \\ \hline
\}                                                                                                     & else \{                                                                       & 5   & 6  & -- &  \\ \hline
\end{tabular}
\end{table*}

In order to obtain these \textit{features} values, the original and generated lines are stripped to remove trailing white spaces. If either of the lines is a combination of code and comment, the code and comment are separated for comparison. The code of the original line is compared with the code of the generated line to calculate the \texttt{ld}. The comment of the original line is compared with the comment of the generated line to generate the \texttt{BLEU} metric. Additionally, if the original line is a comment, the location of the most recent previous comment is updated to the current line, and the \texttt{dfc} is calculated. The lines generated and \textit{features} obtained, for example, in \autoref{fig:llm-inputs-example}, are shown in \autoref{tbl:example-features}. For the example shown, LLM gpt-3.5-turbo was used in auto-complete mode. It is not possible to obtain \texttt{logprob} values for gpt-3.5-turbo through the public API so those values are not shown. \texttt{logprob} values are available for experiments with code-davinci-002.

\subsection{Classification\label{subsec:classification}}

Classification for a given file  takes the \textit{features} as inputs and selects lines to flag based on some conditions. These conditions are referred to as \textit{criteria}, and the lines flagged are referred to as \textit{reported\_lines}. The conditions can be inclusive or exclusive. The inclusive conditions are designed to bring lines into the set of \textit{reported\_lines}. The exclusive conditions are used to eliminate  lines from \textit{reported\_lines}.
The inclusive conditions use two thresholds in our FLAG framework, i.e., Levenshtein Distance Upper Limit (\texttt{ld\_limit}) and Distance from Comment Upper Limit (\texttt{dfc\_limit}). 
 These thresholds are chosen for reasons detailed below.
The \texttt{ld\_limit} and \texttt{dfc\_limit} work together to yield different \textit{criteria}.

Why \texttt{ld\_limit}? \texttt{ld} indicates the difference between two pieces of code. If \texttt{ld} is 0, the pieces of code are identical. This means that the LLM does not have an alternative suggestion, so there is no reason to flag this code. If \texttt{ld} is greater than 0, the pieces of code are different, indicating that the LLM is indicating an alternative to the original line of code. This could be worthy of flagging. But if \texttt{ld} is a very high number, it could indicate that the LLM is generating something completely different. Based on the understanding that a buggy version of code is normally very similar to the fixed version, we use an upper limit to \texttt{ld} to target the generated code that is different but not too different.

Why \texttt{dfc\_limit}? \texttt{dfc} indicates how far a line of code is from the closest comment before it. If there is a comment close to the code, i.e., \texttt{dfc} has a low value, we hypothesize that the LLM would generate code with relevant information, so we trust it more. If \texttt{dfc} is greater than \texttt{dfc\_limit}, the comment is likely not relevant to the line of code, and we discard it. The use of \texttt{dfc} relaxes our criteria for selecting lines to flag as a line of code that has a \texttt{ld} greater than \texttt{ld\_limit} may still be flagged if it has a \texttt{dfc} less than \texttt{dfc\_limit}.

A simple criterion C0 uses the \texttt{ld\_limit} as its threshold.
\begin{equation*}
C0(ld\_limit):0<ld<=ld\_limit
\end{equation*}
A more elaborate criterion C1 uses both thresholds, 
\begin{equation*}
\begin{aligned}
C1&(ld\_limit,dfc\_limit):\\
&0<ld\;AND\;(ld<=ld\_limit\;OR\; 0<dfc<dfc\_limit)
\end{aligned}
\end{equation*}
The final criterion C2 uses both thresholds and employs a \textit{reduce\_fp()} function to remove false positives. %
\begin{equation*}
\begin{aligned}
C2&(ld\_limit,dfc\_limit):\\
&0<ld\;AND\;(ld<=ld\_limit\;OR\; 0<dfc<dfc\_limit)\\
&AND\;reduce\_fp()
\end{aligned}
\end{equation*}

The side-effect of  detecting defects is a substantial number of false positives in \textit{reported\_lines}. We tackle this by adopting a few measures in \textit{reduce\_fp()}. The process inspects \textit{reported\_lines} to remove some of the flagged lines that are likely false positives.
The first measure is recomputing \texttt{ld} after removing white spaces in generated and original lines. This removes false positives where the \texttt{ld} counted white spaces, e.g., \texttt{always(@posedge clk)} and \texttt{always (@posedge clk)}.
The second is checking if the original line has just a keyword. If this is the case, that line is removed from \textit{reported\_lines} because a simple keyword can not be buggy.
The third uses \texttt{logprob} values as thresholds for exclusion. If the LLM has a large negative \texttt{logprob} value for the generated line, it is removed from \textit{reported\_lines} as it indicates that the LLM is not confident in its suggestion. We use a threshold of $<-0.5$ for our experiments with code-davinci-002 to remove lines. Classification of the example shown in~\autoref{fig:llm-inputs-example}  using C2(20,10) is shown in column \texttt{Flagged?} of~\autoref{tbl:example-features}. Of the 2 flagged lines, one contains the original defect (shaded in pink), while the other is a false positive.

\section{Benchmark Dataset\label{sec:dataset}}

To evaluate the feasibility of using LLMs to flag inconsistencies in the comment or code, we experiment on a set of example defects in C, Python, and Verilog from various sources, summarized in \autoref{tbl:sources}. We collect both security-related and functional defects. Sources C1, P1, and V1 consist of security defects, whereas sources C2, P2, and V2 have defects that pose a functional issue. 
Some of the security-related defects are described in~\autoref{tbl:sec-bugs}, and the remaining security and functional defects are described in Appendix~\ref{subsec:bugs-details}. 19 of the 35 security defects have a weakness present in MITRE's top 25 CWEs list. Common Weakness Enumerations (CWE)s are categories of security issues that can manifest in code~\cite{the_mitre_corporation_cwe_2022}.
Each defect has a source code file and corresponding line number(s) that contains the defect. We gather 121 defects; 43 for C, 34 for Python, and 44 for Verilog.

\begin{table*}[h]
\centering
\caption{Sources for defects in C, Python, and Verilog. Func=Functional, Sec=Security}
\label{tbl:sources}
\small
\begin{tabular}{|L{1.5cm}|l|L{0.7cm}|L{11.2cm}|L{0.9cm}|}
\hline
\textbf{Source} & \textbf{Lang.} & \textbf{Type} & \textbf{What we used as defects/bugs} & \textbf{\#Bugs} \\ \hline
C1~\cite{pearce_asleep_2022,pearce_examining_2022}                                                             & C                 & Sec          & Insecure code instances from CVEs and CWEs that authors used to engineer repairs. & 13               \\ \hline
C2~\cite{yi_feasibility_2017}                                                              & C                 & Func         & Defects in students' submissions for introductory programming assignments.                     & 30               \\ \hline
P1~\cite{pearce_asleep_2022,pearce_examining_2022}                                                              & Python            & Sec           & Insecure code instances from CVEs and CWEs that authors used to engineer repairs. & 12               \\ \hline
P2~\cite{widyasari_bugsinpy_2020}                                                              & Python            & Func       & Bugs extrapolated from patch files for a subset of the bugs in this database.                  & 22               \\ \hline
V1~\cite{ahmad_fixing_2023}                                                             & Verilog           & Sec           & Defects introduced in modules from various sources to evaluate their repair tool.          & 10               \\ \hline
V2~\cite{ahmad_cirfix_2022}                                                             & Verilog           & Func         & Defects introduced in common Verilog modules by authors to evaluate their repair tool.      & 34               \\ \hline
\end{tabular}
\end{table*}

\begin{table*}[h]
\centering
\caption{An example selection of security-related bugs and descriptions (The full table is in Appendix~\ref{sec:appendix}). (*) indicates that the CWE is in one of MITRE's top 25 CWEs list.}
\label{tbl:sec-bugs}
\small
\begin{tabular}{|l|l|L{14.5cm}|}
\hline
\textbf{ID} & \textbf{CWE} &\textbf{Description} \\ \hline
C1-1  & 125$^*$ & libjpeg-turbo 2.0.1 has a heap-based buffer over-read in the put\_pixel\_rows function.   \\ \hline
C1-2  & 119$^*$ & Heap-based buffer overflow in libjpeg-turbo 1.2.0 allows remote attackers to cause a denial of service.     \\ \hline
C1-3  & 119$^*$ & The DumpModeDecode function in libtiff 4.0.6 and earlier allows attackers to cause a denial of service.                                                                                                            \\ \hline

P1-1  & 79$^*$ & Writing user input directly to a web page allows for a cross-site scripting vulnerability.                                                                    \\ \hline
P1-2  & 20$^*$ & Security checks on the substrings of an unparsed URL are often vulnerable to bypassing.                                                                                                                                  \\ \hline
P1-3  & 22$^*$ & Uncontrolled data used in path expression can allow an attacker to access unexpected resources.                                \\ \hline
V1-1  & 1234 & Lock protection is overridden when debug mode is active.                                      \\ \hline
V1-2  & 1271 & A locked register does not have a value assigned on reset. When it is brought out of reset, the state is unknown.                                            \\ \hline
V1-3  & 1280 & An asset is allowed to be modified even before the access control check is complete.   \\ \hline
\end{tabular}
\end{table*}

\subsection{C sources\label{sec:c-sources}}

C1 contains security vulnerabilities curated by the authors of~\cite{pearce_asleep_2022,pearce_examining_2022}. We chose 12 real-world CVEs and 5 instances of CWEs investigated in those works. Common Vulnerabilities and Exposures (CVE) is a glossary that classifies vulnerabilities. A particular exposure in the real world is documented and given an identification. We investigated cve-2018-19664, cve-2012-2806, cve-2016-5321, cve-2014-8128, cve-2014-8128, cve-2016-10094, cve-2017-7601, cve-2016-3623, cve-2017-7595, cve-2016-1838, cve-2012-5134 and cve-2017-5969. The details of each CVE could be obtained from the NIST National Vulnerability Database~\cite{national_vulnerability_database_nvd_2023}. For CWEs, we included cwe-119 and cwe-125, which were inspired by  MITRE examples, and cwe-416, cwe-476, cwe-732, taken from CodeQL examples.

C2 contains defects in assignments submitted by students~\cite{yi_feasibility_2017}. There were 74 unique assignments spread across 10 weeks. We selected 30 of these assignments with topics including Simple Expressions, Loops, Integer Arrays, Character Arrays (Strings) and Functions, Multi-dimensional Arrays, Recursion, Pointers, Algorithms (sorting, permutations, puzzles), and Structures (User-Defined data-types). For each of the chosen assignments, we picked a submission by a student which had a clearly identifiable bug. This was done by comparing the buggy version with the correct version in the repository posted by the authors of the work.
Each unique assignment also contains a correct version of the code with comments at the start describing the intended functionality of the code. We use these comments and add them as prefix to the relevant source code files we analyze.

\subsection{Python sources\label{sec:p-sources}}
P1 uses the work by authors to use Github Copilot to complete code based on scenarios developed for select CWEs~\cite{pearce_asleep_2022,pearce_examining_2022}. We look at completions marked as containing the weakness by the authors and selected one for each CWE for our work. We cover cwe-79, cwe-20, cwe-22, cwe-78, cwe-89, cwe-200, cwe-306, cwe-434, cwe-502, cwe-522, cwe-732 and cwe-798.

P2 has defects noted in real-world Python projects by analyzing their version control histories~\cite{widyasari_bugsinpy_2020}. This was done by identifying commits on Github that intend to fix a bug. We chose one or two bugs identified in the youtube-dl, tqdm, fastapi, luigi, scrapy, black, nvbn, spacy, keras, pysnooper, cookiecutter and ansible projects. The defects cover a range of causes, including wrong regex expression,
incorrect returned objects, incorrect split token used, incorrect assignment, incorrect parameter value, incorrect os command, incorrect function argument, incorrect computation, wrong error condition, missing file encoding, incorrect appended element to list, incorrect file path in os command and others. For each of these defects we isolate the source code file containing the bug and its location by analyzing the \textit{bug\_patch.txt} file for the bug.

\subsection{Verilog sources\label{sec:v-sources}}

V1 contains defects in the form of security vulnerabilities gathered by the authors from 3 sources~\cite{ahmad_fixing_2023} . The first source is MITRE's hardware CWEs, from which they design 4 modules with the CWEs present (a locked register, a reset logic for register, an access checker, and a TrustZone peripheral). The second source is the OpenTitan SoC~\cite{lowrisc_contributors_open_2023}, where authors insert bugs in the RTL of 3 modules (ROM Controller, One Time Programmable Memory Controller, and interface for KMAC Key Manager). The third is bugs in Hack@DAC 2021 SoC~\cite{hackevent_hackdac21_2022} that uses the Ariane core with bugs present for the bug-hunting competition. The bugs are in the Control and Status Register Regfile, Direct Memory Access, and AES interface modules.

V2 contains functional defects in 11 Verilog projects~\cite{ahmad_cirfix_2022}. These include a 3-to-8 decoder, 4-bit counter with overflow, t-flip flop, finite state machine, 8-bit left shift register, 4-to-1 multiplexer, i2c bidirectional serial bus, sha3 cryptographic hash function, core for the Tate bilinear pairing algorithm for elliptic curves, Core for Reed Solomon error correction and an SDRAM controller. The defects include numeric errors, incorrect sensitivity lists, incorrect assignments, incorrect resets, blocking instead of non-blocking assignments,  wrong increment of counter and skipped buffer overflow check.

\section{Experiments and Results\label{subsec:tool-experiments-and-results}}

To evaluate FLAG consistency checking approach, we design experiments using 2 OpenAI LLMs.
The first LLM is code-davinci-002 which is optimized for code-completion tasks. For code-davinci-002, there is no system prompt, but it supports an insert capability if a suffix is given in addition to the prefix. We run 2 experiments with code-davinci-002, in 
auto-complete (without suffix) and insertion (with suffix) modes.
The second LLM is gpt-3.5-turbo. It improves on GPT-3 and can understand and generate natural language or code. At the time of this manuscript, it was the "most capable GPT-3.5 model". %
It is an instructional model whose role is deducted based on the system prompt it is given. After the system prompt is given, the model is then given the message which specifies the  task. We conduct two experiments with gpt-3.5-turbo. The first is done without any system prompt. We refer to this mode auto-complete. The second gives the following system prompt: \textit{"You are a skilled AI programming assistant. Complete the next line of code."}. This mode is instructed-complete. 

An experiment involves choosing a particular set of LLM and mode, e.g., code-davinci-002 in insertion mode, and using it to execute our consistency checking approach on all 121 benchmarks. In total, we run 4 experiments for the possible combinations of 2 LLMs and their 2 modes of completion. 
To evaluate the success of an experiment, we focus on 3 metrics. 
The first is the \texttt{Number of defects detected (DD)}. For a given set of inputs, \texttt{DD} is the sum of defects that were correctly identified. It is effectively the True Positives(TP) and has a maximum possible value of 121, i.e., total defects across all sources.
The second is the \texttt{False Positive Rate (FPR)}. For a given set of inputs, \texttt{FPR} is the ratio between the number of lines incorrectly highlighted and the total number of lines.
The third is the \texttt{True Positive Rate (TPR)} or \texttt{Recall} i.e., the ratio between the true positives and total number of positives. This is the \texttt{DD}/total number of defects.

\subsection{Results\label{subsubsec:evaluation-framework}}

The results are presented in \autoref{tbl:results-summary}. We break down the results for each experiment by source and criteria to present a nuanced view. For C, we break down C1 into C1-CVEs and C1-CWEs because of the difference in their size and results by FLAG checker.
By showing snapshots of different criteria, we illustrate how the results change based on complexity. Ultimately, we propose criterion 2 as it balances TPR and FPR and uses inclusive and exclusive \textit{features} from \autoref{subsubsec:features}.

gpt-3.5-turbo has a greater ability to detect defects, but code-davinci-002 has fewer false positives. For criterion 2, gpt-3.5-turbo performed better than code-davinci-002 in terms of \texttt{TPR}. It was able to detect 90 of the defects in auto-complete and 77 in instructed-insertion mode, while code-davinci-002 detected 76 in auto-complete and 78 in insertion. code-davinci-002 performs better in terms of \texttt{FPR}. It has a lower FPR of 0.121 in auto-complete and 0.141 in insertion compared to 0.172 and 0.181 for gpt-3.5-turbo in instructed-complete and auto-complete, respectively.
The criteria take particular values for \texttt{ld\_limit} and \texttt{dfc\_limit}. This need not be the case as they can take a range of values which can impact the results. We discuss these relations and insights of the nature of true and false positives in~\autoref{sec:analysis}.

\begin{table*}[h]
\centering
\footnotesize
\caption{Results Summary. Number of defects detected (DD), False Positive Rate (FPR) and True Positive Rate (TPR) shown for each combination of LLM, Mode of completion, Language, Source and Criteria. C0 has a ld\_limit of 10 i.e C0(10). C1 has ld\_limit of 20 and dfc\_limit of 10 i.e., C1(20,10). C2 has the same limits as C1 but employs the reduce\_fp() function, i.e., C2(20,10). Numbers in bold summarize the metrics of all sources for a mode of completion of an LLM. Bold number in DD column is the sum of DD for all sources. Bold numbers in FPR and TPR columns are the FPR and TPR for all sources.}
\label{tbl:results-summary}

\resizebox{\textwidth}{!}{
\begin{tabular}{lccccrrrrrrrrrr}
\cline{2-15}
                     & \multirow{2}{*}{Model}          & \multirow{2}{*}{\begin{tabular}[c]{@{}c@{}}Mode of \\ Completion\end{tabular}} & \multicolumn{1}{l}{\multirow{2}{*}{Language}} & \multirow{2}{*}{Source} & \multicolumn{1}{l}{}           & \multicolumn{3}{c}{Criterion 0}                                             & \multicolumn{3}{c}{Criterion 1}                                             & \multicolumn{3}{c}{Criterion 2}                                             \\ \cline{7-15} 
\multicolumn{1}{c}{} &                                 &                                                                                & \multicolumn{1}{l}{}                          &                         & \multicolumn{1}{c}{\# Defects} & \multicolumn{1}{c}{DD} & \multicolumn{1}{c}{FPR} & \multicolumn{1}{l}{TPR} & \multicolumn{1}{c}{DD} & \multicolumn{1}{c}{FPR} & \multicolumn{1}{l}{TPR} & \multicolumn{1}{c}{DD} & \multicolumn{1}{c}{FPR} & \multicolumn{1}{l}{TPR} \\ \cline{2-15} 
                     & \multirow{16}{*}{davinci-002}   & \multirow{8}{*}{\begin{tabular}[c]{@{}c@{}}auto\\ complete\end{tabular}}       & \multirow{3}{*}{C}                            & C1- CVEs                & 8                              & 1                      & 0.071                   & 0.125                   & 2                      & 0.113                   & 0.250                   & 2                      & 0.098                   & 0.250                   \\
                     &                                 &                                                                                &                                               & C1-CWEs                 & 5                              & 1                      & 0.101                   & 0.200                   & 2                      & 0.163                   & 0.400                   & 2                      & 0.124                   & 0.400                   \\
                     &                                 &                                                                                &                                               &  C2                      & 30                             & 19                     & 0.088                   & 0.633                   & 25                     & 0.111                   & 0.833                   & 24                     & 0.093                   & 0.800                   \\ \cline{4-15} 
                     &                                 &                                                                                & \multirow{2}{*}{Verilog}                      & V1                      & 10                             & 1                      & 0.081                   & 0.100                   & 7                      & 0.216                   & 0.700                   & 7                      & 0.206                   & 0.700                    \\
                     &                                 &                                                                                &                                               & V2                      & 34                             & 20                     & 0.089                   & 0.588                   & 26                     & 0.172                   & 0.765                   & 25                     & 0.149                   & 0.735                  \\ \cline{4-15} 
                     &                                 &                                                                                & \multirow{2}{*}{Python}                       & P1                      & 12                             & 5                      & 0.063                   & 0.417                   & 10                     & 0.165                   & 0.833                   & 7                      & 0.092                   & 0.583                   \\
                     &                                 &                                                                                &                                               & P2                      & 22                             & 5                      & 0.064                   & 0.227                   & 9                      & 0.163                   & 0.409                   & 9                      & 0.147                   & 0.409                    \\ \cline{4-15} 
\texttt{}            &                                 &                                                                                & \textbf{}                                     & \textbf{}               & \textbf{121}                   & \textbf{52}            & \textbf{0.075}          & \textbf{0.430}          & \textbf{81}            & \textbf{0.138}          & \textbf{0.669}          & \textbf{76}            & \textbf{0.121}          & \textbf{0.628}          \\ \cline{3-15} 
                     &                                 & \multirow{8}{*}{insertion}                                                     & \multirow{3}{*}{C}                            & C1- CVEs                & 8                              & 2                      & 0.143                   & 0.250                   & 2                      & 0.144                   & 0.250                   & 2                      & 0.126                   & 0.250                   \\
                     &                                 &                                                                                &                                               & C1-CWEs                 & 5                              & 1                      & 0.147                   & 0.200                   & 2                      & 0.163                   & 0.400                   & 2                      & 0.124                   & 0.400                   \\
                     &                                 &                                                                                &                                               & C2                      & 30                             & 24                     & 0.108                   & 0.800                   & 25                     & 0.111                   & 0.833                   & 24                     & 0.094                   & 0.800                   \\ \cline{4-15} 
                     &                                 &                                                                                & \multirow{2}{*}{Verilog}                      &V1                      & 10                             & 4                      & 0.155                   & 0.400                   & 7                      & 0.215                   & 0.700                   & 7                      & 0.204                   & 0.700                   \\
                     &                                 &                                                                                &                                               & V2                      & 34                             & 25                     & 0.148                   & 0.735                   & 26                     & 0.174                   & 0.765                   & 25                     & 0.150                   & 0.735                   \\ \cline{4-15} 
                     &                                 &                                                                                & \multirow{2}{*}{Python}                       & P1                      & 12                             & 7                      & 0.209                   & 0.583                   & 9                      & 0.296                   & 0.750                   & 7                      & 0.175                   & 0.583                   \\
                     &                                 &                                                                                &                                               & P2                      & 22                             & 9                      & 0.145                   & 0.409                   & 11                     & 0.196                   & 0.500                   & 11                     & 0.168                   & 0.500                   \\ \cline{4-15} 
                     &                                 &                                                                                &                                               &                         & \textbf{121}                   & \textbf{72}            & \textbf{0.145}          & \textbf{0.595}          & \textbf{82}            & \textbf{0.162}          & \textbf{0.678}          & \textbf{78}            & \textbf{0.141}          & \textbf{0.645}          \\ \cline{2-15} 
                     & \multirow{16}{*}{gpt-3.5-turbo} & \multirow{8}{*}{\begin{tabular}[c]{@{}c@{}}auto\\ complete\end{tabular}}       & \multirow{3}{*}{C}                            &C1- CVEs                & 8                              & 0                      & 0.089                   & 0.000                   & 4                      & 0.165                   & 0.500                   & 4                      & 0.137                   & 0.500                   \\
                     &                                 &                                                                                &                                               & C1-CWEs                 & 5                              & 0                      & 0.147                   & 0.000                   & 3                      & 0.302                   & 0.600                   & 3                      & 0.256                   & 0.600                   \\
                     &                                 &                                                                                &                                               & C2                      & 30                             & 19                     & 0.162                   & 0.633                   & 27                     & 0.245                   & 0.900                   & 25                     & 0.219                   & 0.833                   \\ \cline{4-15} 
                     &                                 &                                                                                & \multirow{2}{*}{Verilog}                      &V1                      & 10                             & 0                      & 0.116                   & 0.000                   & 8                      & 0.352                   & 0.800                   & 8                      & 0.318                   & 0.800                   \\
                     &                                 &                                                                                &                                               & V2                      & 34                             & 18                     & 0.135                   & 0.529                   & 30                     & 0.294                   & 0.882                   & 29                     & 0.251                   & 0.853                  \\ \cline{4-15} 
                     &                                 &                                                                                & \multirow{2}{*}{Python}                       &P1                      & 12                             & 1                      & 0.092                   & 0.083                   & 11                     & 0.282                   & 0.917                   & 11                     & 0.189                   & 0.917                   \\
                     &                                 &                                                                                &                                               & P2                      & 22                             & 4                      & 0.060                   & 0.182                   & 10                     & 0.193                   & 0.455                   & 10                     & 0.176                   & 0.455                   \\ \cline{4-15} 
                     &                                 &                                                                                &                                               &                         & \textbf{121}                   & \textbf{42}            & \textbf{0.099}          & \textbf{0.347}          & \textbf{93}            & \textbf{0.210}          & \textbf{0.769}          & \textbf{90}            & \textbf{0.181}          & \textbf{0.744}          \\ \cline{3-15} 
                     &                                 & \multirow{8}{*}{\begin{tabular}[c]{@{}c@{}}instructed\\ complete\end{tabular}} & \multirow{3}{*}{C}                            & C1- CVEs                & 8                              & 1                      & 0.096                   & 0.125                   & 3                      & 0.168                   & 0.375                   & 3                      & 0.131                   & 0.375                   \\
                     &                                 &                                                                                &                                               & C1-CWEs                 & 5                              & 0                      & 0.124                   & 0.000                   & 3                      & 0.279                   & 0.600                   & 3                      & 0.256                   & 0.600                   \\
                     &                                 &                                                                                &                                               &C2                      & 30                             & 19                     & 0.133                   & 0.633                   & 24                     & 0.224                   & 0.800                   & 22                     & 0.193                   & 0.733                   \\ \cline{4-15} 
                     &                                 &                                                                                & \multirow{2}{*}{Verilog}                      & V1                      & 10                             & 0                      & 0.127                   & 0.000                   & 6                      & 0.372                   & 0.600                   & 6                      & 0.318                   & 0.600                   \\
                     &                                 &                                                                                &                                               & V2                      & 34                             & 14                     & 0.159                   & 0.412                   & 26                     & 0.312                   & 0.765                   & 25                     & 0.257                   & 0.735                   \\ \cline{4-15} 
                     &                                 &                                                                                & \multirow{2}{*}{Python}                       &P1                      & 12                             & 2                      & 0.102                   & 0.167                   & 10                     & 0.282                   & 0.833                   & 10                     & 0.194                   & 0.833                   \\
                     &                                 &                                                                                &                                               &  P2                      & 22                             & 3                      & 0.055                   & 0.136                   & 8                      & 0.160                   & 0.364                   & 8                      & 0.144                   & 0.364                   \\ \cline{4-15} 
\textbf{}            &                                 &                                                                                & \textbf{}                                     & \textbf{}               & \textbf{121}                   & \textbf{39}            & \textbf{0.105}          & \textbf{0.322}          & \textbf{80}            & \textbf{0.210}          & \textbf{0.661}          & \textbf{77}            & \textbf{0.172}          & \textbf{0.636}          \\ \cline{2-15} 
\end{tabular}
}
\end{table*}

\subsection{Deeper Dive into Security Bugs}

\begin{figure}[!tbp]
\centering
\small
\begin{tabular}{|l|l|l|l|l|l|}
\hline

\cellcolor{pink}C1-1  & C1-2  & \cellcolor{pink}C1-3 & C1-4  & \cellcolor{green} C1-5  & \cellcolor{green} C1-6  \\

       &   \color{pink} x \color{pink} x \color{green}\checkmark \color{pink} x    & &   \color{pink} x \color{pink} x \color{green}\checkmark \color{green}\checkmark   &       &          \\
      
      \hline

\cellcolor{pink}C1-7 & \cellcolor{pink}C1-8  & \cellcolor{green} C1-9  & \cellcolor{green} C1-10 & \cellcolor{pink}C1-11 & C1-12 \\

&       &       &       &   &  \color{pink} x \color{pink} x \color{green}\checkmark \color{green}\checkmark             \\

    \hline

   \cellcolor{green} C1-13 & P1-1 & P1-2  & \cellcolor{green} P1-3  & P1-4  & \cellcolor{green} P1-5 \\

   &     \color{pink} x \color{pink} x \color{green}\checkmark \color{green}\checkmark & \color{green}\checkmark \color{green}\checkmark \color{green}\checkmark \color{pink} x   &       &  \color{pink} x \color{pink} x \color{pink} x \color{green}\checkmark     &       \\
   
   \hline

   \cellcolor{green} P1-6  & \cellcolor{green} P1-7  & P1-8  & \cellcolor{green} P1-9  & P1-10 & \cellcolor{green} P1-11 \\
   
           &       &   \color{pink} x \color{pink} x \color{green}\checkmark \color{pink} x    &       &   \color{pink} x \color{pink} x \color{green}\checkmark \color{green}\checkmark    &      \\

           \hline

   P1-12 & V1-1  & \cellcolor{green}  V1-2  & V1-3  & \cellcolor{green} V1-4  & \cellcolor{green} V1-5 \\
   
   \color{pink} x \color{pink} x \color{green}\checkmark \color{green}\checkmark & \color{pink} x \color{pink} x 
   \color{green}\checkmark \color{pink} x    &       &   \color{green}\checkmark \color{green}\checkmark \color{pink} x \color{pink} x    &       &       \\

   \hline

  \cellcolor{green} V1-6  & V1-7  & \cellcolor{green} V1-8  & V1-9  & \cellcolor{pink} V1-10 \\

         &   \color{green}\checkmark \color{green}\checkmark \color{green}\checkmark \color{pink} x     &       &   \color{pink} x \color{pink} x \color{green}\checkmark \color{green}\checkmark   &    \\

         \cline{1-5}
\end{tabular}
\caption{Security bugs detected according to modes and LLMs. A green cell indicates that  LLMs in both modes were able to detect this bug. Conversely, pink indicates that no LLM in any mode was able to detect this defect. The remaining cells have a set of 4 symbols to represent whether the defect was detected by [code-davinci-002 in auto-complete, code-davinci-002 in insertion, gpt-3.5-turbo in auto-complete and gpt-3.5-turbo in instructed-complete]. A sequence of [\color{green}\checkmark\color{green}\checkmark\color{green}\checkmark \color{pink} x \color{black}] means that the defect was detected by code-davinci-002 in both modes and by gpt-3.5-turbo in auto-complete but not in instructed-complete e.g. P1-2.  \label{tbl:details-security-detection}
}

\end{figure}

As we are interested in the security applications of FLAG, we breakdown the detection of security bugs in~\autoref{tbl:details-security-detection}. The breakdown for functional defects is shown in Appendix~\ref{subsec:bugs-detection-details}. 16 of the 35 security bugs were detected by both LLMs in both modes, and 29 were detected by at least one LLM in one mode.
gpt-3.5-turbo in auto-complete mode performed the best, detecting 26 compared to 22, 19 and 19 for gpt-3.5-turbo in instructed-complete, code-davinci-002 in auto-complete and code-davinci-002 in insertion respectively.
Python security bugs were the ones best detected by the LLMs. All defects for Python were identified. It is followed by Verilog and C, respectively. A closer inspection allows us to develop some insights into why this might be the case. Firstly, the size of P1 source files was smaller in comparison to other sources. It had an average file size of 17 lines compared to 1426 of C1 and 249 of V1. Additionally, the small file size allowed the LLM to take in the entire source code before the bug as a part of the prompt, providing it with the complete context of the intended functionality of the files. Secondly, benchmarks in P1 were particularly designed by Pearce et al.~\cite{pearce_asleep_2022,pearce_examining_2022} for security evaluation whereas benchmarks in V1 and C1 were a combination of those explicitly designed for security and real-world examples. 6 of the benchmarks in V1 were from code in real implementations of SoCs, and 8 benchmarks in C1 are from real-world CVEs. This also explains why FLAG performed poorly on C1-CVEs, i.e., C1-1 - C1-8.

\begin{figure}[h]
    \centering
    \includegraphics[width=\linewidth]{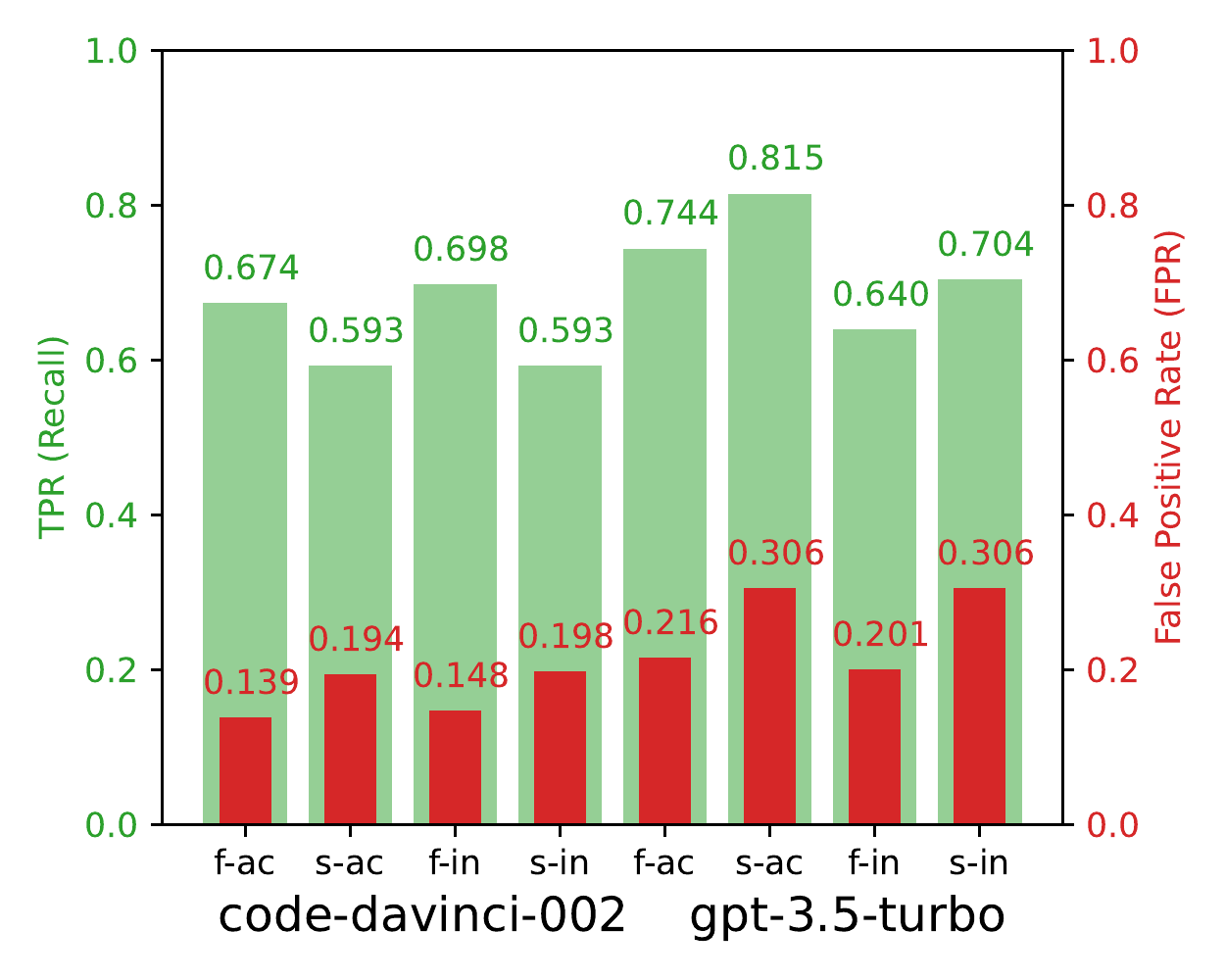}
    \caption{Performance for function and security categories using criterion 2 C2(20,10). `f'=functional, `s'=security, `ac'=auto-complete, `in'=insert for code-davinci-002, `in'=instruct-complete for gpt-3.5-turbo. C2-CVEs excluded.}
    \label{fig:functional-security-comparison}
\end{figure}

We are interested in the comparison of the performance of FLAG between functional and security bugs. Based on data in~\autoref{fig:functional-security-comparison}, one may deduce that FLAG detects functional bugs better than security bugs. Considering results at C2(20,10), functional bugs have a higher average TPR of 0.689 compared to that for security bugs at 0.676. They also have a lower average FPR of 0.176 compared to that for security bugs at 0.271. A closer look reveals that Python is an exception with a higher TPR for P1 as compared to P2. For gpt-3.5-turbo the TPR for P1 is twice that of P2, indicating that gpt-3.5-turbo does well in detecting security bugs in Python. We exclude data for C1-CVEs from the  analysis  because the number of lines for these benchmarks is many times larger than others, skewing averages towards  C1-CVEs.

\subsection{Detailed Analysis\label{sec:analysis}}

\begin{figure}[!b]
    \centering
    \includegraphics[width=\linewidth]{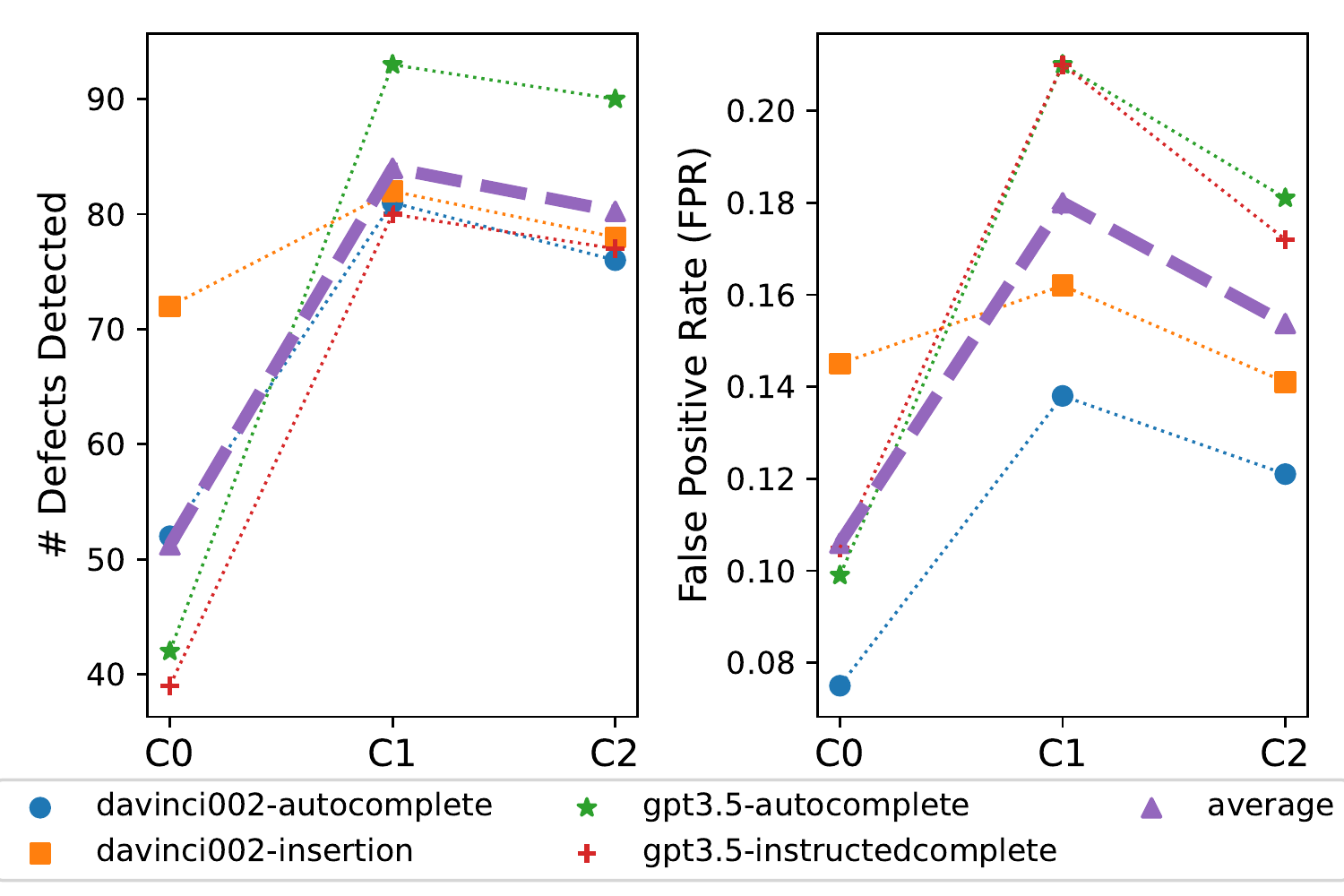}
    \caption{Trend of \texttt{DD} (left) and \texttt{FPR} (right) for C0 - C2 criteria. Combined trend for LLMs and modes is the average in purple.}
    \label{fig:trends_across_criteria}
\end{figure}

\textbf{How does refining criteria from C0 to C2 impact DD and AFPR?}
Moving from C0 to C1, \texttt{DD} and \texttt{FPR} both increase. This is because \texttt{ld\_limit} is considerably relaxed from 10 to 20, and \texttt{dfc\_limit} is used to include lines that may exceed the \texttt{ld\_limit}. On average, the \texttt{DD} increases from 51 to 84 while the \texttt{FPR} increases from 0.106 to 0.18. 
Moving from criteria C1 to C2, \texttt{DD} and \texttt{FPR} both decrease. This is because the \textit{reduce\_fp()} function removes highlighted lines from \textit{reported\_lines}. Some false positives were removed, but so were some true positives. On average, the \texttt{DD} decreases from 85 to 80 while the \texttt{FPR} decreases from 0.18 to 0.154. While the \texttt{DD} decreases by 4.76\%, the \texttt{FPR} decreases by 14.6\%. This signifies importance of \textit{reduce\_fp()} in improving the performance of consistency checking.
The trends are shown in \autoref{fig:trends_across_criteria}.

\begin{figure}[!tbp]
    \centering
    \includegraphics[width=\linewidth]{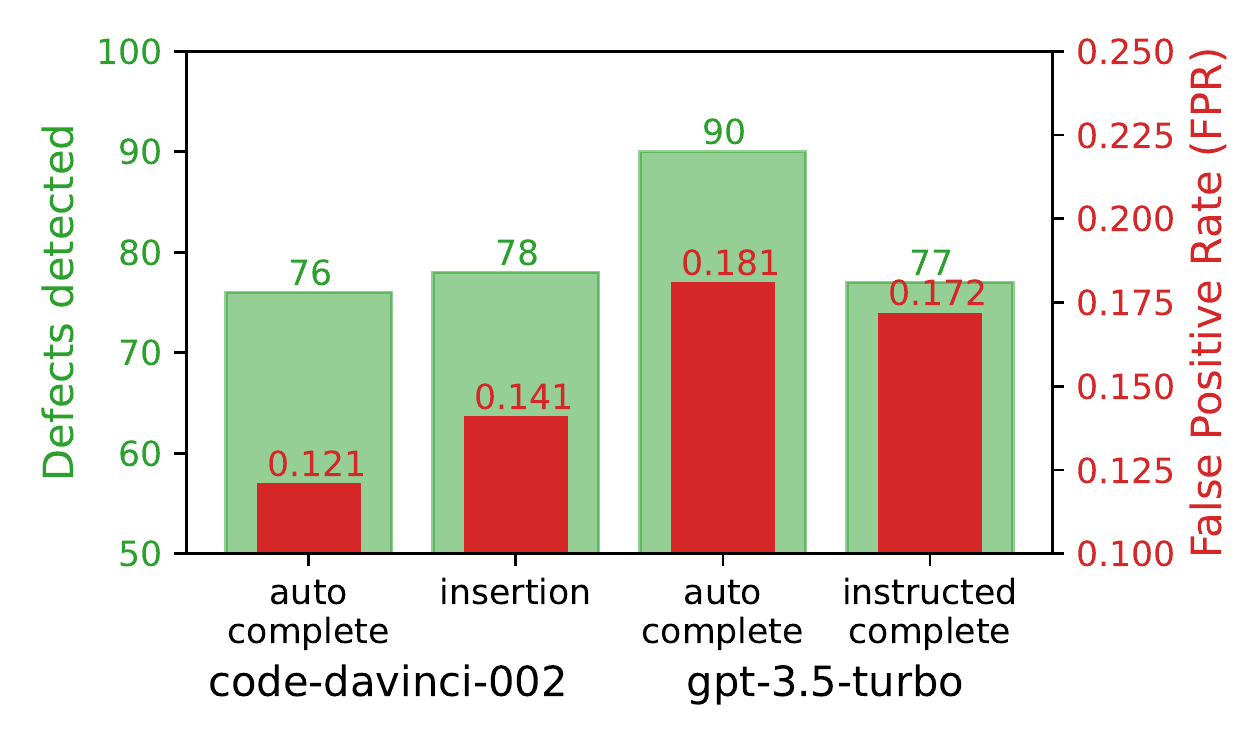}
    \caption{Performance of completion modes for C2(20,10).}
    \label{fig:completion-mode-comparison}
\end{figure}

\textbf{How do the different completion modes impact DD and AFPR?}
Instinctively, we may think that an insertion mode should do better than the auto-complete mode, as the LLM has access to more information in the form of the suffix, in addition to the prefix. Additionally, instructing gpt-3.5-turbo to produce "the next line of code" in the instructed-complete mode should perform better than the auto-complete mode as the likelihood of generating code instead of an explanation of code should be higher. 
\autoref{fig:completion-mode-comparison} compares the performances of different completion modes at C2(20,10). 
For code-davinci, the insertion mode does allow us to detect 2 additional defects but at the cost of 16.5\% increase in \texttt{FPR}. This is probably not a valuable enough trade-off, meaning that not a lot of benefit is obtained from using the insertion mode.
For gpt-3.5-turbo, the instructed-complete mode does significantly worse than its auto-complete counterpart. It detects 13 fewer defects while decreasing the \texttt{FPR} by only 4.97\%. 

\begin{figure}[!b]
    \centering
    \begin{subfigure}[b]{0.46\linewidth}
    \includegraphics[height=5cm,width=\linewidth]{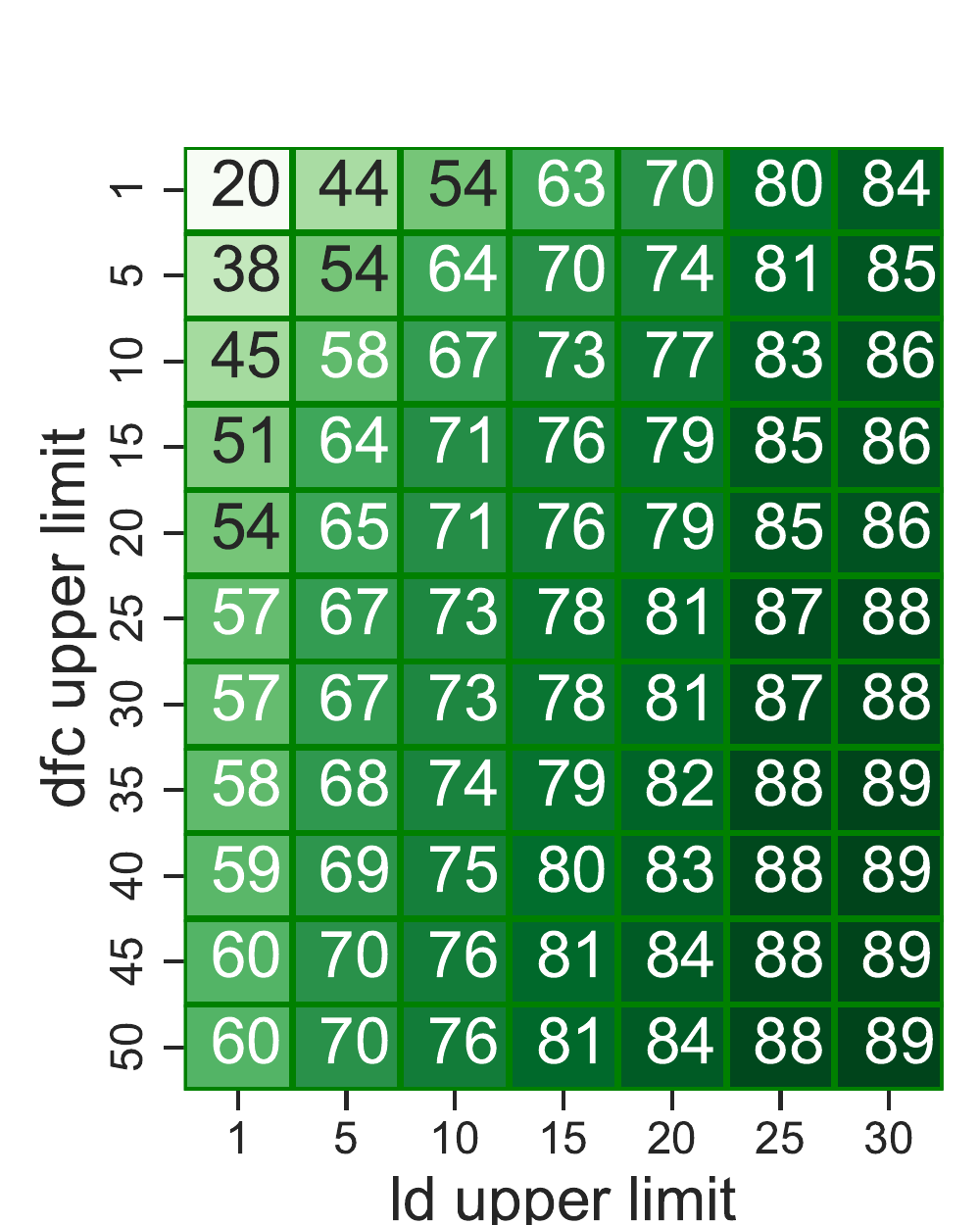}
    \caption{Defects detected (DD)}
    \label{fig:defects-heatmap}
    \end{subfigure}
    \hfill
    \begin{subfigure}[b]{0.53\linewidth}
    \includegraphics[height=5cm,width=\linewidth]{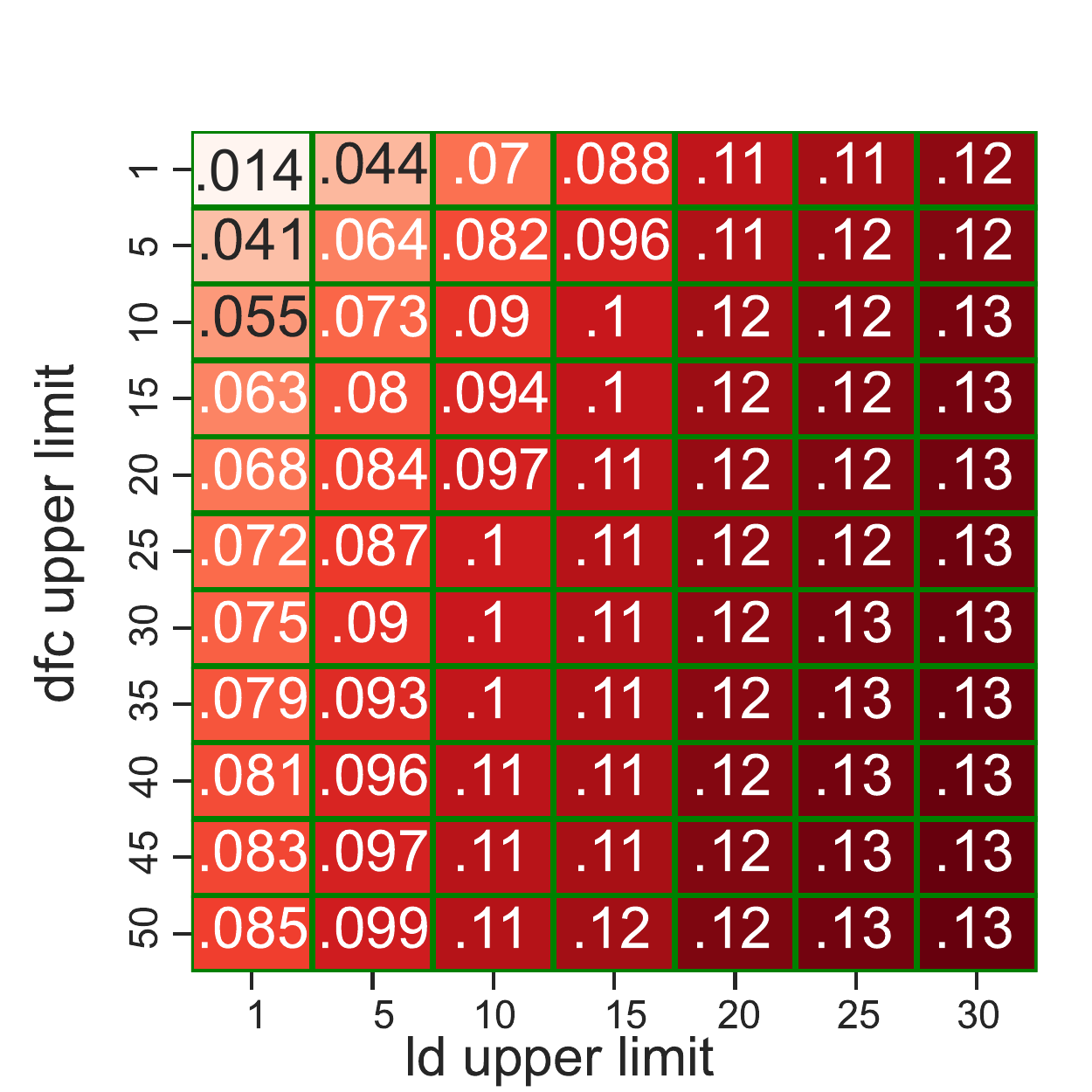}
    \caption{False Positive Rates (FPR)s}
    \label{fig:false-positives-heatmap}
    \end{subfigure}
    
    \caption{Defects, FPRs for (ld\_limit,dfc\_limit) combinations for code-davinci-002 auto-complete mode.}
\end{figure}

\textbf{How do \texttt{ld\_limit} and \texttt{dfc\_limit} impact DD and AFPR?} The two main determinants of highlighting lines in our checking approach are \texttt{ld\_limit} and \texttt{dfc\_limit}. Since they can take any continuous value, there is a need for a deeper analysis other than considering the particular values discussed in criteria C0 to C2. 
We conduct sweeps across the two limits for the code-davinci-002
model in auto-complete mode to see their impacts on the \texttt{DD} and \texttt{FPR}.
\autoref{fig:defects-heatmap} and \autoref{fig:false-positives-heatmap} show how the \texttt{DD} and \texttt{FPR} change respectively at different combinations of \texttt{ld\_limit} and \texttt{dfc\_limit} in the range from 0 to 30 for \texttt{ld\_limit} and 0 to 50 for \texttt{dfc\_limit}.
We observe that the \texttt{ld\_upper\_limit} is much more dominant in impacting the \texttt{DD} and \texttt{FPR}. A change in \texttt{ld\_upper\_limit} brings about a greater change in both \texttt{DD} and \texttt{FPR} compared to an equal change in \texttt{dfc\_limit}. As a result, \texttt{DD} and \texttt{FPR} also saturate earlier while increasing \texttt{ld\_limit}. \texttt{DD} loosely saturates  at \texttt{ld\_limit} of 30 as there is limited impact of changing \texttt{dfc\_limit} at that value. Similarly, \texttt{FPR} loosely saturates at \texttt{ld\_limit} of 20. These heatmaps are partial, but extending them for \texttt{ld\_limit}=50 reveals that the maximum possible values for \texttt{DD} and \texttt{FPR} are 90 and 0.14, respectively.
We also observe that \texttt{DD} and \texttt{FPR} are more sensitive at smaller values of \texttt{ld\_limit} and \texttt{dfc\_limit}. We can achieve 80\% of the maximum value of \texttt{DD}, i.e., 72 at \texttt{ld\_limit} of 15 and \texttt{dfc\_limit} of 10. Similarly, we can achieve 80\% of the maximum value of \texttt{FPR} i.e., 0.11 at \texttt{ld\_limit} of 15 and \texttt{dfc\_limit} of 25. This analysis helps in gaining an estimate of what are the ranges that should be considered when designing similar tools.

\begin{figure}[!tbp]
    \centering
    \includegraphics[width=\linewidth]{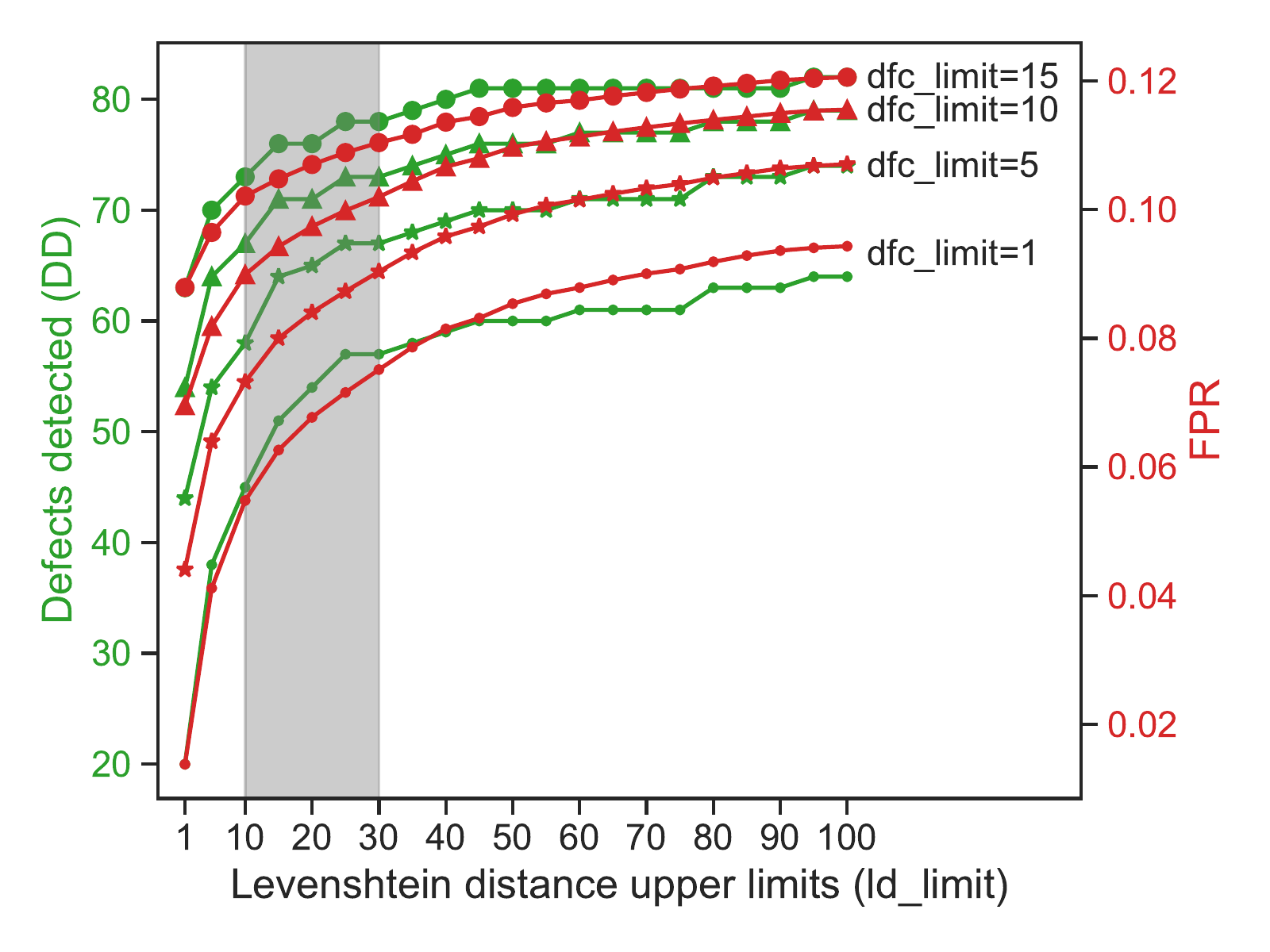}
    \caption{Change in Defects Detected and False Positive Rate with Levenshtein distance upper limit (ld\_limit). Highlighted region produces decent results for  \texttt{dfc\_limit} values.}
    \label{fig:progression_with_ld}
\end{figure}

This can be explored further by zooming into the smaller values for \texttt{dfc\_limit} and extending the values for \texttt{ld\_limit} to 100 as shown in~\autoref{fig:progression_with_ld}. For a particular \texttt{dfc\_limit}, as we increase the \texttt{ld\_limit}, both \texttt{DD} and \texttt{FPR} increase. While the rate of increase for both decreases (shown by the decreasing gradient), there is a point after which the rate of increase in \texttt{DD} is significantly less than that of \texttt{FPR}. This is the point beyond which the benefit of obtaining more defects would be overpowered by the cost of increasing false positives. Conversely, a much smaller value of \texttt{ld\_limit} would simply not find enough defects. The highlighted region in~\autoref{fig:progression_with_ld} indicates the values for which the consistency checking approach provides ``decent'' results i.e., between 10 and 30. 
A similar exploration with keeping \texttt{ld\_limit} constant and sweeping \texttt{dfc\_limit} provides similar insights. The values for \texttt{DD} and \texttt{FPR}, however, are not distinct enough for different values of \texttt{ld\_limit} for a graphical representation to be illustrative.

Naturally, a question may be asked of what is the optimal combination of \texttt{ld\_limit} and \texttt{dfc\_limit}?
How many false positives a coder is willing to accept for the benefit of detecting more bugs is a matter of subjectivity. We illustrated the trade-off between the \texttt{TPR} and the \texttt{FPR} in the heatmaps. A way of deciding which limits to use could be by first looking at the number of false positives you are ready to bear. For example, if you are willing to look at 7 lines out of 100 for every bug detected, i.e., \texttt{FPR} of 0.07, you may choose \texttt{(ld\_limit,dfc\_limit)} of (5,10) as shown in \autoref{fig:false-positives-heatmap}. At these limits, you would be able to locate 58 of the 121 defects covered in our benchmarks as that is the corresponding value in \autoref{fig:defects-heatmap}.

\begin{figure}[!tbp]
    \centering
    \begin{subfigure}[b]{0.45\linewidth}
    \centering
    \includegraphics[width=\linewidth]{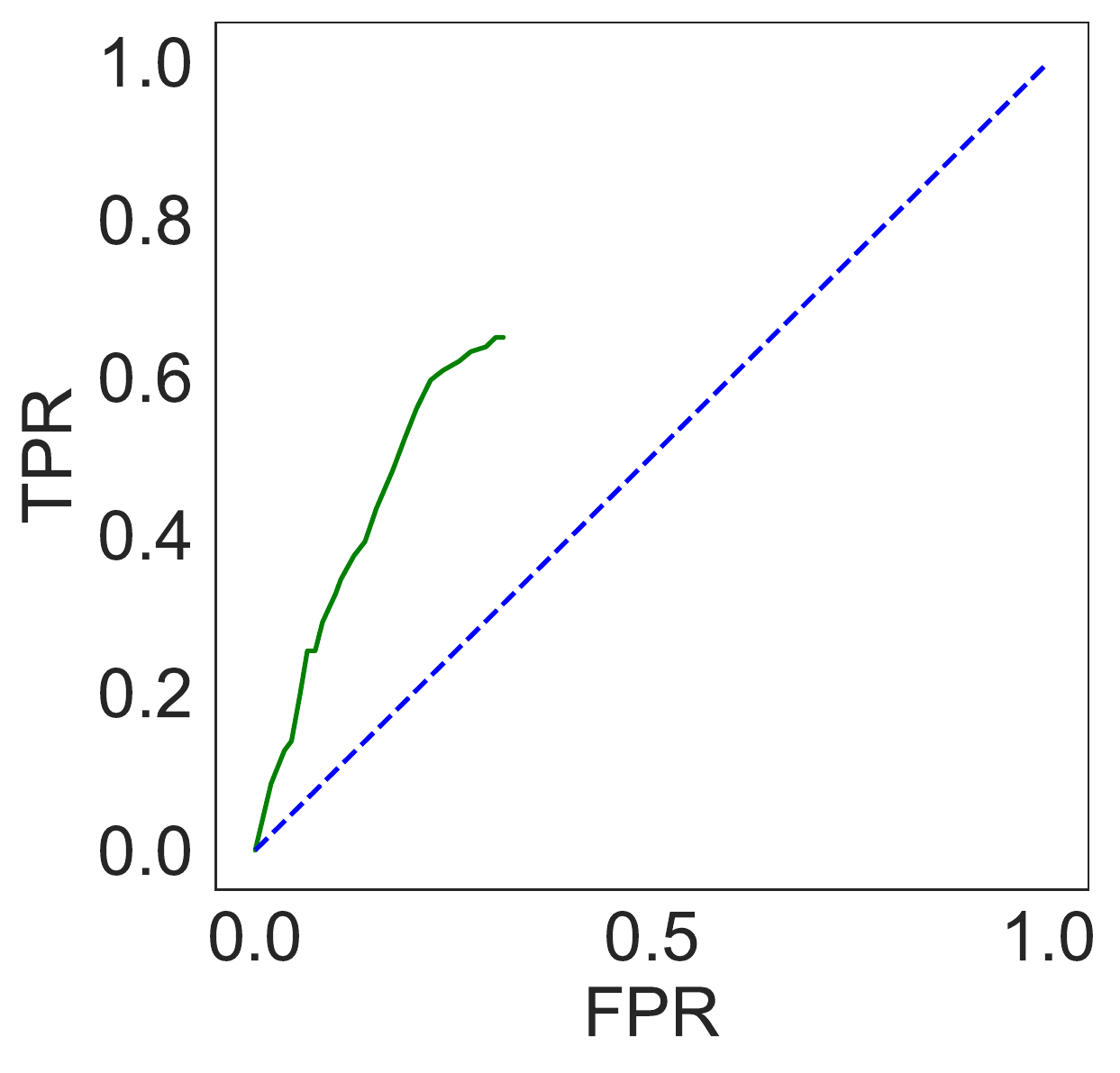}
    \caption{\texttt{ld} lower limit is set to 0.}
    \label{fig:roc-partial}
    \end{subfigure}
    \hfill
    \begin{subfigure}[b]{0.45\linewidth}
    \centering
    \includegraphics[width=\linewidth]{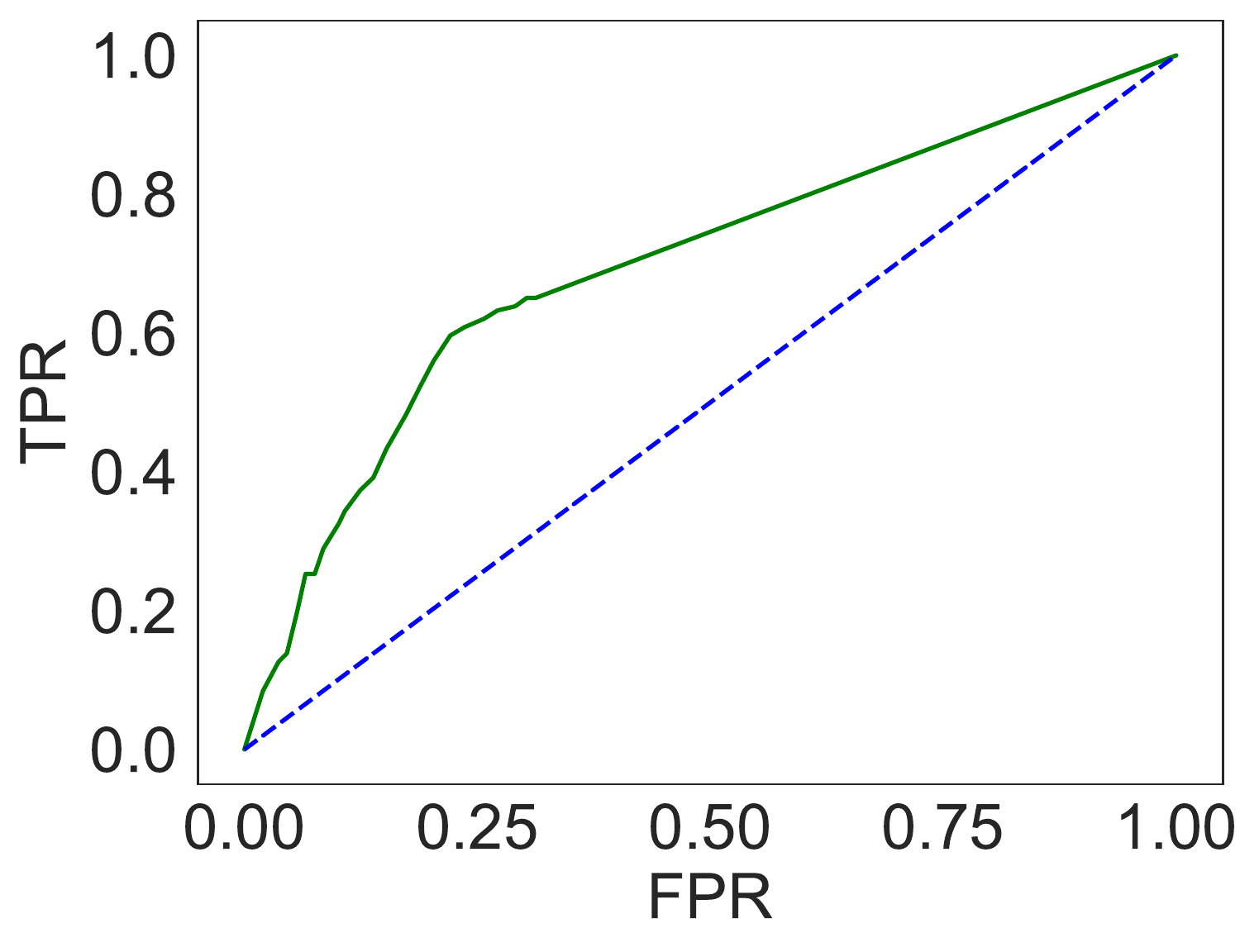}
    \caption{Point added for \texttt{ld} lower limit at -1 and upper limit set at 1000.}
    \label{fig:roc}
    \end{subfigure}
    
    \caption{Receiver Operating Characteristic (ROC) when \texttt{ld\_limit} are threshold for code-davinci002 auto-completes.}
\end{figure}

\textbf{How much better are LLMs in relation to random guess?} In order to illustrate the success of consistency checking as a classifier, we represent the \texttt{TPR} and \texttt{FPR} in the form of a Receiver Operating Characteristic (ROC) curve at different thresholds for \texttt{ld\_limit}. We used code-davinci002 in auto-complete mode for this purpose.
In our experiments, we set a hard lower limit for \texttt{ld} of 0. This means that some of the lines that have a \texttt{ld} of zero will never be counted as a true positive or a false positive, resulting in an incomplete ROC curve. This situation is represented in \autoref{fig:roc-partial}, where the \texttt{TPR} and \texttt{FPR} are capped. The ROC curve, however, lies above the \texttt{TPR==FPR} line showing that our classification methodology is better than guesswork. 
For completeness' sake, if we utilize a lower limit at -1 and upper limit at 1000, we can obtain \texttt{TPR} and \texttt{FPR} of 1 because all lines will be highlighted as containing a defect. This is represented in \autoref{fig:roc}.

\begin{figure}[!tbp]
    \centering
    \includegraphics[width=\linewidth]{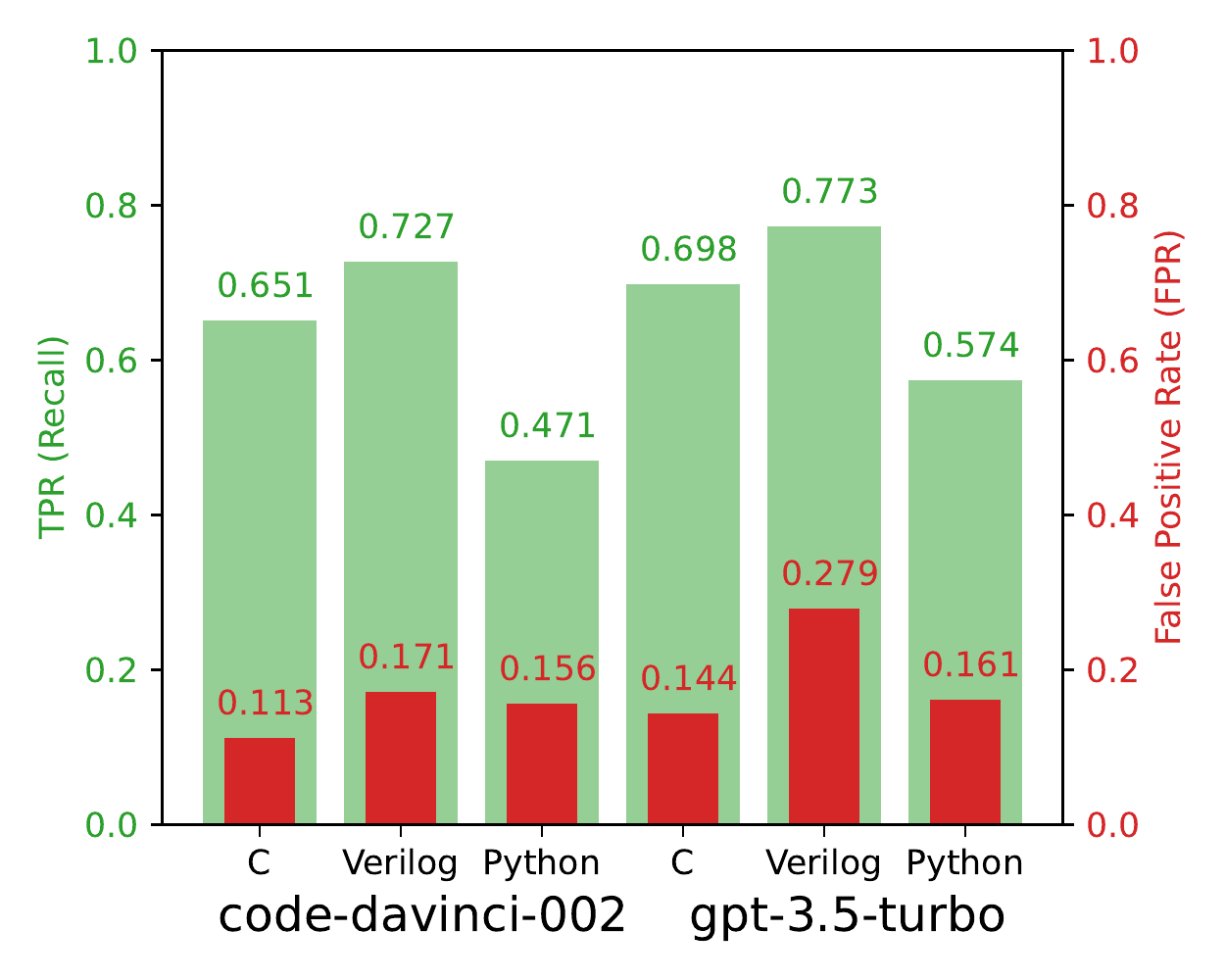}
    \caption{Performance on languages for criterion 2 i.e., C2(20,10). Values represent averages across modes for LLM.}
    \label{fig:languages-comparison}
\end{figure}

\textbf{Are bugs of one language easier to detect than the others?}
Based on the data in \autoref{fig:languages-comparison}, we can conclude that FLAG consistency checker performed the best on C and worst on Python. Although C has a slightly lower \texttt{TPR} than Verilog for both LLMs, it has a significantly lower \texttt{FPR}. While its \texttt{TPR} is 10.5\% lower, its \texttt{FPR} is 48.4\% less than that of Verilog. Python has the lowest \texttt{TPR} and only the second lowest \texttt{FPR} for both LLMs. 
This is surprising because we expect the larger amount of open-source Python in the training data %
would translate to better performance than on Verilog. 
This is probably because 22 of the benchmarks for Python were real-world examples compared to only 6 for Verilog.

\begin{figure}[!tbp]
    \centering
    \includegraphics[width=\linewidth]{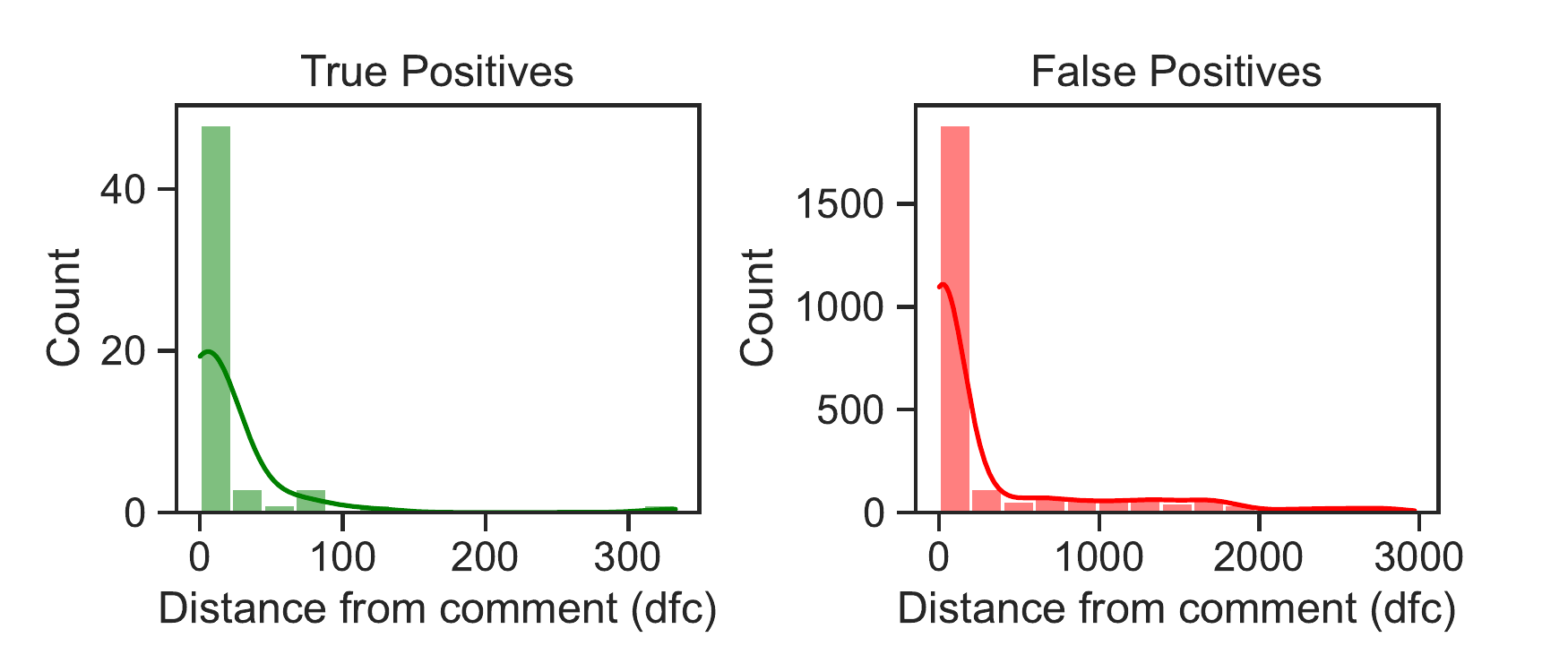}
    \caption{Distance from comment (dfc) distributions for code-davinci-002 in auto-complete mode.}
    \label{fig:dfc-distributions}
\end{figure}

\textbf{How much do comments help?}
We signify the role of comments by analyzing the \texttt{dfc} and \texttt{BLEU} scores for true and false positives. The \texttt{dfc} signifies the closeness of a line to a comment, while the \texttt{BLEU} score signifies the quality of the comment produced by the LLM. 
\autoref{fig:dfc-distributions} shows how the \texttt{dfc} differs between true and false positives. The data here does not include instances where \texttt{dfc} is not available, i.e., when there is no comment before the line concerned. Most of the data in both cases lies in the range for smallest values of \texttt{dfc}. This is because comments are frequently written in benchmarks we study.
\texttt{dfc} for true positives has a lower average of 18.4 compared to that of false positives at 327.9. LLM does a better job at classification when the line concerned is closer to a comment. \texttt{dfc} for true positives  has a much smaller standard deviation of 47.7 compared to that of false positives at 632.1. Thus false positives cover a much larger range of values.

\begin{figure}[h]
    \centering
    \includegraphics[width=\linewidth]{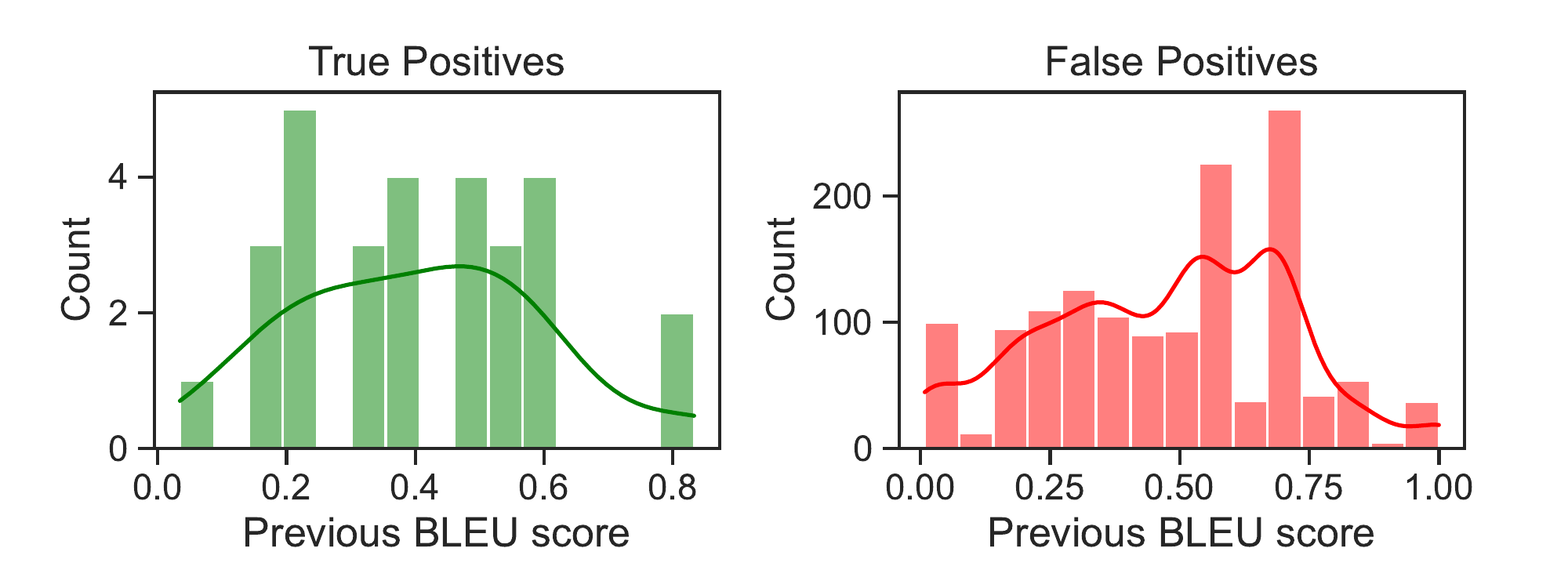}
    \caption{Previous comment's BLEU score distributions for code-davinci-002 in auto-complete mode.}
    \label{fig:prev-BLEU-distributions}
\end{figure}

{\bf Is the average dfc or avg bleu score correlated to the success of the checker?}
\autoref{fig:prev-BLEU-distributions} shows how the previous comment's BLEU score differs between true and false positives. The data does not include lines for which there was no comment before the line or when the LLM was not able to produce a comment. For a line of code being analyzed, FLAG tracks the most recent comment preceding it. FLAG compares the comment produced by the LLM at this line  to the original comment to compute  previous comment's \texttt{BLEU} score. A high value indicates that the LLM is on the right track following the intention of the coder. 
The previous comment's \texttt{BLEU} score for true positives has a similar average of 0.407 compared to that of false positives at 0.473. This shows that the previous comment's \texttt{BLEU} score does not play a crucial role in the classification. The previous comment's \texttt{BLEU} score for true positives  has a similar standard deviation of 0.193 compared to that of false positives at 0.236.

\section{Discussion\label{sec:discussion}}

\begin{figure}[!b]
    \centering
    \includegraphics[width=\linewidth]{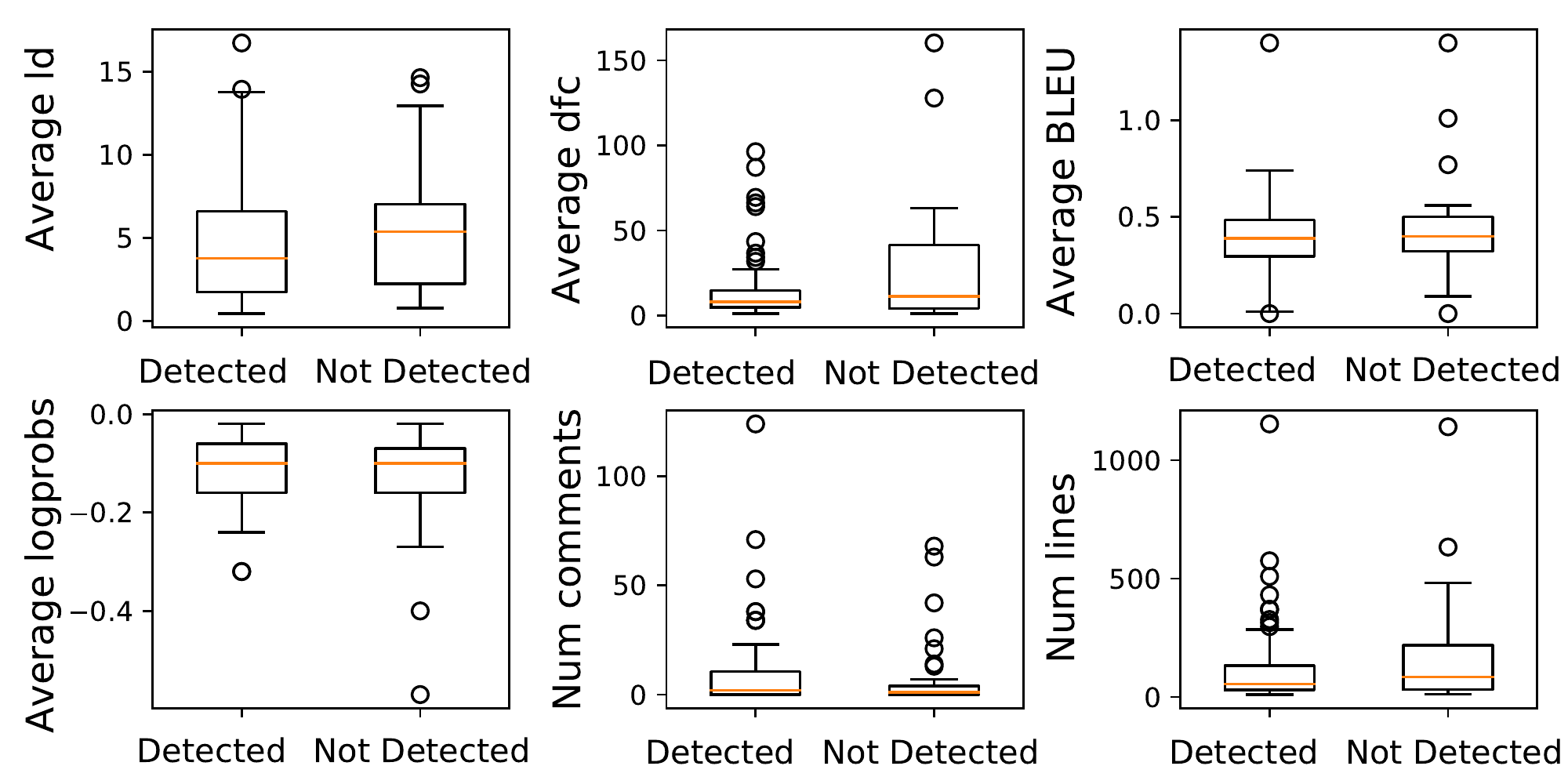}
    \caption{Metadata relation to success of checker. Data used for code-davinci-002 in auto-complete mode for C2(20,10).}
    \label{fig:metadata}
\end{figure}

In the previous section, we explored whether there is a significant difference between true positives and false positives based on information like \texttt{dfc} or previous comment's \texttt{BLEU} scores. This analysis was done at the level of granularity of lines. 
There is however room for such an analysis at the level of granularity of the file or benchmark. Is there a relation between properties of files, like the number of comments, size of benchmark, etc., to the likelihood of a defect being detected for that benchmark? We investigate this relation for the following properties of each benchmark: average \texttt{ld}, average \texttt{dfc}, average \texttt{BLEU}, average \texttt{logprobs}, number of comments and number of lines (size). The results are shown in~\autoref{fig:metadata}. The boxplots reveal that average \texttt{dfc}, number of comments, and number of lines have a significantly different distribution for benchmarks where defects were detected than those where they were not. A lower average \texttt{dfc}, higher number of comments, and smaller number of lines increase the likelihood of the defect being detected. Average \texttt{ld}, average \texttt{BLEU}, and average \texttt{logprobs}, however, do not pose this difference.

FLAG relies on the fundamental idea that a line containing a defect is likely to be written in an alternative way by an LLM i.e., \texttt{ld}$>0$. In our experiments, however, this was not always the case. There were 11 benchmarks for which \texttt{ld} was $0$  for both LLMs in both their modes for C2(20,10), as the LLMs generated the exact same code as the original code. 8 of them were from source V2, 2 from C1 and 1 from P2. Thus, LLMs sometime produce functionally buggy code for Verilog but generally generate alternate code when facing a defect.

A limitation of FLAG is that going line by line for almost every line in the source code does not scale in terms of time. While smaller files $\sim$100 lines are checked in under a minute, scanning a file with thousands of lines can take about an hour. To address this, one could prioritize generating a subset of lines. %
This could be done by checking security-sensitive regions of source code, e.g., checking if conditions and statements inside the if block. Another approach is to ignore lines without much substance, e.g., a line with only a keyword.

Another limitation is the high  \texttt{FPR}. The \texttt{FPR} for our experiments ranges from 0.121 to 0.172, meaning that for a source code file of 100 lines with one defect, about 12-17 lines would be falsely highlighted. %
Future work will focus on further reducing the flagged lines. 
An approach is to consider \textit{features} for regions of source code instead of lines for the criteria for classification, e.g., do not highlight defects between lines x and y for file z because average \texttt{logprobs} for this region indicates that the LLM is not confident in its suggestions.

FLAG can aid a designer by focusing the search for bugs on only $12-17\%$ of the source. Moreover, efforts to reduce false positives should be met with caution as they may eliminate some true positives. FLAG's approach to removing false positives of moving from criteria C1 to C2 reduced the number of defects detected by an average of 3.75. An example is shown in~\autoref{tbl:example-eliminating-true-positive}. The defect is the line highlighted in pink for which the LLM generates alternative code that has an \texttt{ld} of 10 and a \texttt{dfc} of 3. It is highlighted by FLAG as a concern when using C1(20,10) but not when using C2(20,10). This is because C2 uses \textit{reduce\_fp()}, which removes a defective line  as it has a \texttt{logprob} $<-0.5$. FLAG is promising  for bug detection as LLMs improve. With larger data sizes and  fine-tuning, newer LLMs can produce better code and code suggestions.

\begin{table}[t!]
\centering
\caption{Lines generated and features extracted by code-davinci-002 in auto-complete mode for CWE-79 in  P2. Defect in last line was detected with C1 but not C2. logp=logprob.
}
\label{tbl:example-eliminating-true-positive}
\small
\begin{tabular}{|L{2.3cm}|L{2.5cm}|c|c|c|}
\hline
\textbf{Original line}                                                                & \textbf{Generated line}                               & \textbf{dfc} & \textbf{ld} & \textbf{logp} \\ \hline
bp = Blueprint(`hello' \_\_name\_\_)                                                  & bp = Blueprint(`hello'  \_\_name\_\_)                 & 10                                & 1                                & -0.33                                 \\ \hline
\#function which says hello to the user                                               & \#function to run the command                         & 0                                 &           & -1.67                                 \\ \hline
@bp.route("hello/\textless{}username\textgreater{}")                                  & @bp.route(`/hello /\textless{}username\textgreater{}') & 1                                 & 3                                & -0.12                                 \\ \hline
def hello(username):                                                                  & def hello(username):                                  & 2                                 & 0                                & -0.14                                 \\ \hline
\rowcolor[HTML]{FFCCC9} 
return "\textless{}p\textgreater{}Hello  \%s!\textless{}/p\textgreater{}" \% username & return "Hello "+username+"!"                          & 3                                 & 18                               & -0.6                                  \\ \hline
\end{tabular}
\end{table}

\section{Conclusions and Future work\label{sec:conclusion}}

Overall, 101 out of 121 defects were detected by at least one mode of the 2 LLMs. On average, for the 4 combinations of mode and LLMs, 80 defects were detected at an \texttt{FPR} of 0.154 using criterion C2. This gives credence to consistency checks by  FLAG as a bug detector and search space localizer. We find that gpt-3.5-turbo has a better ability to detect bugs but a higher \texttt{FPR} than code-davinci-002. FLAG is marginally better on functional bugs than on security-related bugs. We find that Levenshtein distance between the original and LLM-generated code is the dominant classification feature among those we explored. That said, comments play an important role in the performance of FLAG as lines of code with smaller distances from comments are classified with better success. Lastly, FLAG does best on C and worst on Python for the benchmarks we studied.

Since FLAG is novel in its implementation, several future work directions can be formulated. 
More features in code and LLMs can be used in classification e.g., BLEU scores for code and embedding scores for  code and comments.
With a bigger set of classification features and a  bigger set of true and false positives in the form of benchmarks, ML classifiers could be trained to determine the criteria for classification. Future versions of FLAG could analyze code and comments in chunks rather than  lines. 
Another interesting idea is to run multiple LLMs and flag a line based on the union of features. 

\section*{Acknowledgments}
This research work is supported in part by a gift from Intel Corporation. 
This work does not in any way constitute an Intel endorsement of a product or supplier. 
We acknowledge the support of the Natural Sciences and Engineering Research Council of Canada (NSERC), RGPIN-2022-03027.

\section*{Availability}

The artifacts (FLAG source code, LLM outputs) produced and presented in this study are at~\cite{review_artifacts_2023}. %

\bibliographystyle{plain}
\bibliography{lit/Usenix-LLM-Consistencychecks}
\balance
\setcounter{section}{0}
\renewcommand\thesection{\Alph{section}}
\renewcommand\thesubsection{\thesection.\arabic{subsection}}

\section{Appendix\label{sec:appendix}}

 \label{subsec:bugs-details} \label{subsec:bugs-detection-details}

\autoref{tbl:sec-bugs-full} is the complete table for security-related bugs examined in this work. 
The description for the functional defect benchmarks used and their corresponding IDs are presented in~\autoref{tab:func-benchmarks}.
In~\autoref{tbl:details-detection}, we present the breakdown of the combinations of LLM and their modes that were able to detect each of the collected defects.

\begin{table*}[h!]
\centering
\caption{Complete descriptions for security-related bugs. (*) indicates that the CWE is in one of MITRE's top 25 CWEs list.}
\label{tbl:sec-bugs-full}
\small
\begin{tabular}{|l|l|L{14.5cm}|}
\hline
\textbf{ID} & \textbf{CWE} &\textbf{Description} \\ \hline
C1-1  & 125$^*$ & libjpeg-turbo 2.0.1 has a heap-based buffer over-read in the put\_pixel\_rows function.   \\ \hline
C1-2  & 119$^*$ & Heap-based buffer overflow in libjpeg-turbo 1.2.0 allows remote attackers to cause a denial of service.     \\ \hline
C1-3  & 119$^*$ & The DumpModeDecode function in libtiff 4.0.6 and earlier allows attackers to cause a denial of service.                                                                                                            \\ \hline
C1-4  & 787$^*$ & Segmentation fault due to an out-of-bounds write can be triggered with a malformed TIFF file.                                                                            \\ \hline
C1-5  & 787$^*$ & Out of bounds write in function used to set pixel which may lead to corruption of data, a crash, or code execution.  \\ \hline

C1-6  & 189 & Off-by-one error in LibTIFF 4.0.7 allows remote attackers to have unspecified impact via a crafted image                                                                                              \\ \hline
C1-7  & 20$^*$ & Signed integer overflow in tif\_jpeg.c might allow remote attackers to cause a denial of service. \\ \hline

C1-8  & 369 & rgb2ycbcr tool in LibTIFF 4.0.6 and earlier allows remote attackers to cause a denial of service (divide-by-zero).                                                                                                  \\ \hline
C1-9  & 119$^*$ & Improper restriction of bounds when accessing the memory array. No check for array accessed.                                                                                                                                                                                                 \\ \hline
C1-10 & 125$^*$ & Potential buffer under-read. No check for lower bond of array index.                                                                                                                                                                                                                                \\ \hline
C1-11 & 416$^*$ & An allocated memory block is used after it has been freed. This can cause memory corruption.                                                                                                                                      \\ \hline
C1-12 & 476$^*$ & Dereferencing an untested value from a function that can return null may lead to undefined behavior.                                                                                                                                                                  \\ \hline
C1-13 & 732 & Creating a file that is world-writable can allow an attacker to write to the file.                                                                                                                                                                                   \\ \hline
P1-1  & 79$^*$ & Writing user input directly to a web page allows for a cross-site scripting vulnerability.                                                                    \\ \hline
P1-2  & 20$^*$ & Security checks on the substrings of an unparsed URL are often vulnerable to bypassing.                                                                                                                                  \\ \hline
P1-3  & 22$^*$ & Uncontrolled data used in path expression can allow an attacker to access unexpected resources.                                \\ \hline
P1-4  & 78$^*$ & Externally controlled strings in command line may allow malicious user to change the meaning of the command.       \\ \hline
P1-5  & 89$^*$ & Building a SQL query from user-controlled sources is vulnerable to insertion of malicious SQL code by the user. \\ \hline
P1-6  & 200 & Different messages for an incorrect username, versus a correct username is correct but wrong password.                                                \\ \hline
P1-7  & 306$^*$ & No authentication for functionality that requires a user identity or consumes a significant amount of resources.                                                                                                                                                                                                      \\ \hline
P1-8  & 434$^*$ & Allows the attacker to upload or transfer files of dangerous types that can be automatically processed.                                                                                                                                                                                           \\ \hline
P1-9  & 502$^*$ & The product deserializes untrusted data without sufficiently verifying that the resulting data will be valid.                                                                                                   \\ \hline
P1-10 & 552 & Insufficiently protected credentials make files or directories accessible to unauthorized actors.                                                                                                                                                                                                              \\ \hline
P1-11 & 732 & Overly permissive file permissions allow files to be readable or writable by users other than the owner.                                 \\ \hline
P1-12 & 798$^*$ & Password comparison with a string literal allows an attacker to bypass the authentication that has been configured. \\

\hline
V1-1  & 1234 & Lock protection is overridden when debug mode is active.                                      \\ \hline
V1-2  & 1271 & A locked register does not have a value assigned on reset. When it is brought out of reset, the state is unknown.                                            \\ \hline
V1-3  & 1280 & An asset is allowed to be modified even before the access control check is complete.   \\ \hline
V1-4  & 1276 & The signal depicting the security level for a peripheral instantiated in a SoC is incorrectly grounded. \\ \hline
V1-5  & 1245 & An alert is triggered in the FSM when start signal is high in a state other than Waiting.     \\ \hline
V1-6  & 1245 & The error signal for an FSM moving into invalid state upon global escalation is not asserted.              \\ \hline
V1-7  & 1245 & The done for an FSM is asserted outside of expected window, i.e., during a transmission state.   \\ \hline
V1-8  & 1234 & The core is incorrectly unstalled if there is a request to enter debug mode.                   \\ \hline
V1-9  & 1271 & The register controlling whether the Physical Memory Protection register is writable is not assigned a reset value.       \\ \hline
V1-10 & 1245 & A 4 bit FSM has only 15 states defined and no default statement, resulting in an incomplete case statement.       \\ \hline
\end{tabular}
\end{table*}

\begin{table*}[t]
\centering
\caption{Descriptions for bugs in source C2, P2, and V2}
\label{tab:func-benchmarks}
\scriptsize
\begin{tabular}{@{}|L{0.65cm}|L{4.5cm}|L{0.65cm}|L{4.5cm}|L{0.65cm}|L{4.5cm}|@{}}
\hline
ID    & Description                                   & ID    & Description                                                         & ID    & Description                                                                                                         \\ \hline
C2-1  & Rounding error                                & C2-30 & Decrement instead of increment                                      & V2-7  & Negated if-condition                                                                                                \\ \hline
C2-2  & Incorrect formula                             & P2-1  & wrong regex expression                                              & V2-8  & True and false branches of if-statement swapped                                                                     \\ \hline
C2-3  & compares character instead of integer         & P2-2  & wrong regex expression                                              & V2-9  & Default in case statement omitted                                                                                   \\ \hline
C2-4  & Incorrect calculation                         & P2-3  & incorrect returned objects                                          & V2-10 & Assignment to next state omitted, default cases   in case statement omitted                                         \\ \hline
C2-5  & Using wrong variable                          & P2-4  & incorrect regex expression                                          & V2-11 & state omitted from senslist                                                                                         \\ \hline
C2-6  & Using wrong variable                          & P2-5  & sudo mode not considered                                            & V2-12 & Blocking instead of nonblocking assignments                                                                         \\ \hline
C2-7  & Incorrect condition for prime                 & P2-6  & incorrect split token used                                          & V2-13 & Blocking instead of nonblocking assignments                                                                         \\ \hline
C2-8  & Incorrect condition for prime                 & P2-7  & incorrect assignment                                                & V2-14 & Negated if-condition                                                                                                \\ \hline
C2-9  & incorrect for loop condition                  & P2-8  & incorrect implementation of adding error codes   to string messages & V2-15 & Incorrect sens                                                                                                      \\ \hline
C2-10 & incorrect for loop initializer                & P2-9  & incrrect buffer assignment                                          & V2-16 & Three numeric errors                                                                                                \\ \hline
C2-11 & Incorrect assignment                          & P2-10 & incorrect parameter value                                           & V2-17 & Hex instead of binary numbers                                                                                       \\ \hline
C2-12 & Incorrect array index                         & P2-11 & incorrect os command                                                & V2-18 & 1 bit instead of 4 bit output wire                                                                                  \\ \hline
C2-13 & Incorrect swap                                & P2-12 & incomplete return                                                   & V2-19 & Incorrect sens                                                                                                      \\ \hline
C2-14 & Incorrect string shift amount                 & P2-13 & incorrect function argument                                         & V2-20 & Incorrect assignment                                                                                                \\ \hline
C2-15 & Incorrect input scan                          & P2-14 & incorrect computation                                               & V2-21 & Removed cmd\_ack                                                                                                    \\ \hline
C2-16 & Compare w/ incorrect variable in condition    & P2-15 & wrong error condition                                               & V2-22 & For-loop going too long                                                                                             \\ \hline
C2-17 & No space in print                             & P2-16 & incorrect assignment                                                & V2-23 & Incorrect assignment to out\_ready                                                                                  \\ \hline
C2-18 & Incorrect input scanning                      & P2-17 & missing file encoding                                               & V2-24 & Not checking for buffer overflow during   assignment                                                                \\ \hline
C2-19 & Recursion not returned                        & P2-18 & incomplete function arguments                                       & V2-25 & Logical instead of bitwise negation                                                                                 \\ \hline
C2-20 & Incorrect comparison                          & P2-19 & incorrect implementation of identifying column                      & V2-26 & Incorrect bitshifting                                                                                               \\ \hline
C2-21 & Incorrect condition                           & P2-20 & incorrect appended element to list                                  & V2-27 & Incorrect instantiation of module                                                                                   \\ \hline
C2-22 & Missing statement                             & P2-21 & incorrect condition                                                 & V2-28 & \textbackslash{}textgreater instead of   \textbackslash{}textgreater\{\}\textbackslash{}textgreater for bitshifting \\ \hline
C2-23 & Incorrect increment                           & P2-22 & incorrect file path in os command                                   & V2-29 & Insufficient bits for numeric value                                                                                 \\ \hline
C2-24 & Program not returned                          & V2-1  & Numeric error                                                       & V2-30 & Removed @posedge reset from senslist (sync vs   async reset)                                                        \\ \hline
C2-25 & Incorrect while loop condition                & V2-2  & Numeric error                                                       & V2-31 & wr\_data\_r not reset correctly                                                                                     \\ \hline
C2-26 & Incorrect sorting condition                   & V2-3  & Incorrect assignment                                                & V2-32 & rd\_data\_r assigned incorrectly                                                                                    \\ \hline
C2-27 & Statements wrongly excluded from if condition & V2-4  & Incorrect sens                                                      & V2-33 & Numeric error in parameter                                                                                          \\ \hline
C2-28 & Incorrect assignments to object attributes    & V2-5  & Else-if instead of if                                               & V2-34 & Default in case statement omitted                                                                                   \\ \hline
C2-29 & Wrong formula                                 & V2-6  & Counter never reset                                                 &       &                                                                                                                     \\ \hline
\end{tabular}
\end{table*}

\begin{figure*}[h]
\centering
\begin{tabular}{|l|l|l|l|l|l|l|l|l|l|l|}
\hline

\cellcolor{pink} C2-1  & \cellcolor{green} C2-2  & \cellcolor{green} C2-3  & C2-4  & \cellcolor{pink} C2-5  & \cellcolor{green} C2-6  & \cellcolor{green} C2-7  &  C2-8  & \cellcolor{green} C2-9  & \cellcolor{green} C2-10 & \cellcolor{green} C2-11 \\
  &    &       &    \color{pink} x \color{pink} x \color{green}\checkmark \color{pink} x  &       &       &       &   \color{pink} x \color{pink} x \color{green}\checkmark\color{green}\checkmark   &     &       &       \\
      \hline
      
\cellcolor{green} C2-12 & C2-13 & \cellcolor{green} C2-14 & \cellcolor{green} C2-15 & C2-16 & \cellcolor{green} C2-17 & \cellcolor{green} C2-18 & \cellcolor{green} C2-19 & \cellcolor{green} C2-20 & \cellcolor{green} C2-21 & \cellcolor{green} C2-22 \\
      &  \color{pink} x \color{pink} x \color{green}\checkmark\color{green}\checkmark      &       &       &    \color{green}\checkmark\color{green}\checkmark\color{green}\checkmark \color{pink} x   &       &       &       &       &       &       \\
      \hline
      
\cellcolor{green} C2-23 & C2-24 & \cellcolor{pink} C2-25 & \cellcolor{green} C2-26 & \cellcolor{green} C2-27 & C2-28 & \cellcolor{green} C2-29 & C2-30  &  \cellcolor{green} P2-1 & P2-2  & \cellcolor{pink}P2-3  \\
      &   \color{green}\checkmark\color{green}\checkmark \color{pink} x \color{pink} x    &       &       &       &   \color{green}\checkmark\color{green}\checkmark \color{pink} x \color{pink} x    &       &    \color{green}\checkmark\color{green}\checkmark\color{green}\checkmark \color{pink} x &   \color{pink} x \color{green}\checkmark \color{green}\checkmark \color{green}\checkmark    &  \color{pink} x \color{green}\checkmark \color{green}\checkmark \color{green}\checkmark    &            \\
      \hline

 \cellcolor{green} P2-4  & \cellcolor{pink}P2-5  & \cellcolor{pink}P2-6  & \cellcolor{pink}P2-7  & \cellcolor{green} P2-8  & P2-9  & \cellcolor{green} P2-10 & P2-11 & \cellcolor{pink}P2-12 & P2-13 & P2-14  \\
 &       &       &       &       &       &   \color{pink} x \color{pink} x \color{green}\checkmark \color{green}\checkmark    &       &    \color{green}\checkmark \color{green}\checkmark \color{pink} x \color{green}\checkmark   &   \color{green}\checkmark \color{pink} x \color{green}\checkmark \color{pink} x     &    \color{green}\checkmark \color{pink} x \color{pink} x \color{pink} x      \\
      \hline
      
 P2-15 & \cellcolor{pink}P2-16 & P2-17 & \cellcolor{pink}P2-18 & P2-19 & \cellcolor{pink}P2-20 & P2-21 & P2-22 & \cellcolor{green} V2-1  & \cellcolor{green} V2-2  & V2-3   \\
   \color{pink} x \color{green}\checkmark \color{green}\checkmark \color{green}\checkmark    &       &   \color{green}\checkmark \color{green}\checkmark \color{green}\checkmark \color{pink} x    &       &    \color{pink} x \color{green}\checkmark \color{green}\checkmark \color{pink} x   &       &    \color{pink} x \color{green}\checkmark \color{pink} x \color{pink} x   &    \color{green}\checkmark \color{green}\checkmark \color{pink} x \color{pink} x  &       &      & \color{green}\checkmark \color{green}\checkmark \color{green}\checkmark \color{pink} x       \\
      \hline

 \cellcolor{pink}V2-4  & \cellcolor{green} V2-5  &\cellcolor{green}  V2-6  & \cellcolor{green} V2-7  & \cellcolor{green} V2-8  & \cellcolor{pink} V2-9  & \cellcolor{green} V2-10 & \cellcolor{pink} V2-11 & V2-12 & \cellcolor{green} V2-13 & \cellcolor{green} V2-14  \\
        &       &       &       &       &       &       &    & \color{pink} x \color{pink} x \color{green}\checkmark \color{green}\checkmark    &       &           \\
      \hline
      
 \cellcolor{green} V2-15 & V2-16 & V2-17 & \cellcolor{green} V2-18 & V2-19 & \cellcolor{green} V2-20 & V2-21 & V2-22 & V2-23 & \cellcolor{green} V2-24 & V2-25 \\       &   \color{pink} x \color{pink} x \color{green}\checkmark \color{green}\checkmark    &   \color{pink} x \color{pink} x \color{green}\checkmark\color{green}\checkmark    &       &  \color{pink} x \color{pink} x \color{green}\checkmark \color{green}\checkmark     &       &   \color{pink} x \color{pink} x \color{green}\checkmark \color{pink} x    &  \color{green}\checkmark \color{green}\checkmark \color{green}\checkmark \color{pink} x  & \color{green}\checkmark \color{green}\checkmark \color{pink} x \color{pink} x    &       &   \color{green}\checkmark \color{green}\checkmark \color{pink} x \color{pink} x      \\
      \hline
      
 \cellcolor{green} V2-26 & \cellcolor{green} V2-27 & \cellcolor{green} V2-28 & \cellcolor{green} V2-29 & \cellcolor{green} V2-30 & \cellcolor{green} V2-31 & \cellcolor{green} V2-32 & \cellcolor{green} V2-33 & V2-34 \\
   &       &       &       &       &       &       &    &
  \color{pink} x \color{pink} x \color{green}\checkmark \color{pink} x    \\
      \cline{1-9}

\end{tabular}
\caption{ Functional bugs detected according to mode and LLM. A cell highlighted in green indicates that both LLMs in both modes were able to detect this defect. A cell highlighted in pink indicates that no LLM in any mode was able to detect this defect. The remaining cells present a sequence of 4 symbols to represent whether the defect was detected by code-davinci-002 in auto-complete, code-davinci-002 in insertion, gpt-3.5-turbo in auto-complete and gpt-3.5-turbo in instructed-complete respectively. A sequence of \color{green}\checkmark\color{green}\checkmark\color{green}\checkmark \color{pink} x \color{black} means that the defect was detected by code-davinci-002 in both modes and by gpt-3.5-turbo in auto-complete but not in instructed-complete e.g. C2-16. \color{pink} x \color{pink} x \color{pink} x \color{green}\checkmark \color{black} means that the defect was detected only by gpt-3.5-turbo in instructed-complete.  \label{tbl:details-detection}}

\end{figure*}

\end{document}